\documentclass[twocolumn,bbm,twocolappendix]{aastex631}
\usepackage{graphicx}
\usepackage{natbib}
\usepackage{amsmath}
\usepackage{mathtools}
\usepackage{amssymb}
\usepackage{wasysym}
\usepackage{verbatim}
\usepackage{txfonts}
\usepackage[T1]{fontenc}
\usepackage{ae, aecompl}
\usepackage{hyperref}
\usepackage{upgreek}
\usepackage{xcolor}
\usepackage{latexsym,longtable,epsf,ulem}
\usepackage{cases} 
\usepackage{enumerate} 
\newcommand{\EQ}{\begin{equation}}
\newcommand{\EN}{\end{equation}}
\newcommand{\EQA}{\begin{eqnarray}}
\newcommand{\ENA}{\end{eqnarray}}

\newcommand{\pf}{p_{\rm f}}
\newcommand{\PI}{PI}

\newcommand{\ted}{{t_{\rm ed}}}

\newcommand{\lmag}{\ell_{M}}
\newcommand{\lf}{{\ell_{\rm f}}}
\newcommand{\kf}{{k_{\rm f}}}
\newcommand{\urms}{{u_{\rm rms}}}

\newcommand{\kk}{\mbox{\boldmath $k$} {}}

\newcommand{\FF}{\mbox{\boldmath $F$} {}}

\newcommand{\B}{{\bm B}}

\newcommand{\pc}{\, {\rm pc}}

\newcommand{\ghz}{\,{\rm GHz}}

\def\Rey{{\rm Re}}
\def\Pm{\rm Pm}
\newcommand{\meanF}{{\langle F\rangle}}

\newcommand{\pmean}{{\langle p_{\rm f} \rangle}}
\newcommand{\ncc}{{N_{\rm CC}}}
\newcommand{\ppeak}{{p_{\rm peak}}}
\newcommand{\bra}[1]{\langle #1\rangle}

\shorttitle{Morphology of polarized emission}
\shortauthors{Dutta, Sur, \& Basu}
\received{2024 March 26}
\revised{2024 October 17}
\accepted{2024 October 17}

\begin{document}

\title{Probing the Morphology of Polarized Emission Induced by Fluctuation Dynamo Using Minkowski Functionals}

\correspondingauthor{Sharanya Sur}
\email{sharanya.sur@iiap.res.in}

\author[0009-0000-5591-9063]{Riju Dutta}
\affiliation{Department of Physics, Indian Institute of Science, Bangalore 560012, India}

\author[0000-0003-4286-8476]{Sharanya Sur}
\affiliation{Indian Institute of Astrophysics, 2nd Block, Koramangala, Bangalore 560034, India}

\author[0000-0003-2030-3394]{Aritra Basu}
\affiliation{Th\"uringer Landessternwarte, Sternwarte 5, D-07778 Tautenburg, Germany}
\affiliation{Max-Planck-Institut f{\"u}r Radioastronomie, Auf dem H{\"u}gel 69, D-53121 Bonn, Germany}

\begin{abstract}
The morphology and the characteristic scale of polarized structures provide crucial insights 
into the mechanisms that drive turbulence and maintain magnetic fields in magneto-ionic 
plasma. We aim to establish the efficacy of Minkowski functionals as quantitative statistical 
probes of filamentary morphology of polarized synchrotron emission resulting from fluctuation 
dynamo action. Using synthetic observations generated from magnetohydrodynamic simulations 
of fluctuation dynamos with varying driving scales ($\lf$) of turbulence in isothermal, 
incompressible, and subsonic media, we study the relation between different morphological 
measures and their connection to fractional polarization ($\pf$). We find that Faraday 
depolarization at low frequencies gives rise to small-scale polarized structures that have higher 
filamentarity as compared to the intrinsic structures that are comparable to $\lf$. 
Above $\sim3\ghz$, the number of connected polarized structures per unit area ($N_{\rm CC, peak}$) 
is related to the mean $\pf$ ($\pmean$) of the emitting region as $\pmean \propto N_{\rm CC, peak}^{-1/4}$, 
provided the scale of the detectable emitting region is larger than $\lf$. This implies that 
$N_{\rm CC,peak}$ represents the number of turbulent cells projected on the plane of 
the sky and can be directly used to infer $\lf$ via the relation $\lf \propto N_{\rm CC,peak}^{-1/2}$. 
An estimate of $\lf$ thus directly allows for pinning down the turbulence-driving mechanism 
in astrophysical systems. While the simulated conditions are mostly prevalent in the intracluster 
medium of galaxy clusters, the qualitative morphological features are also applicable in the 
context of interstellar medium in galaxies.
\end{abstract}


\section{Introduction}

Magnetic fields are pervasive in almost all astrophysical objects in our Universe 
and are believed to be amplified and maintained by some form of dynamo action 
\citep[see][for reviews]{C14, SS21}. Polarization observations of magnetic fields 
in turbulent media probe the 2D projection of the ordered and random component 
of the fields in the plane of the sky. Along with information on Faraday depth that 
probes the magnetic field component parallel to the line of sight (LOS), rudimentary 
insight into the 3D structure of magnetic fields can be obtained under certain 
assumptions \citep{beck15}. Such measurements are often challenging to interpret 
\citep{basu19b}, and therefore, directly connecting them to the intrinsic magnetic field 
morphology and its 3D statistical properties is rather arduous.

In the majority of astrophysical systems, where the bulk of the diffuse media are 
turbulent, fast, and efficient, \textit{fluctuation dynamos} are believed to amplify 
and maintain magnetic fields \citep{K68, R19, SS21}. These dynamos are capable 
of amplifying weak, initial seed magnetic fields embedded in a conducting fluid to 
near-equipartition strengths by 3D turbulent flows
\citep[e.g.,][]{Scheko+04,HBD04,Cho+09,BS13,Fed16,Seta+20,XL16,XL21,SS24}.
The morphology of magnetic fields generated by these dynamos
and the associated emission, e.g., via the synchrotron mechanism, are 
intermittent, strongly non-Gaussian, wherein the characteristic emission scale 
depends on the driving scale of turbulence \citep{SBS21,BS21}.
It is therefore imperative to quantify the nature of polarized structures in order to 
glean information on the underlying mechanism of magnetic field amplification.

In recent times, intensity gradient techniques \citep{gaens11, Hu2019, Carmo2020, Wang2021} 
and statistical correlation functions \citep{haver08, mao15, Nandakumar2020, Seta2023} 
have been used to infer the statistical properties of turbulent magneto-ionic media. 
While intensity gradient techniques provide insights into the alignment of the magnetic 
field with various observational tracers, quantitative interpretations often rely on 
benchmarking against numerical simulations. On the other hand, statistical 
correlation functions are mostly insensitive to the morphology of the structures \citep{Seta2018}.
In this work, we study the 2D morphological properties of synthetic polarized 
synchrotron emission produced by fluctuation dynamos driven on different scales 
using quantitative morphological measures based on \textit{Minkowski functionals} 
\citep{AT07}. This technique provides morphological description of any 
structure in $N$-dimensions and has been extensively used to study the 
morphological features in cosmological large-scale structures \citep[e.g.,][]{MBW94,SSS98, bharadwaj00},
anisotropy in maps of the cosmic microwave background \citep{SG98}, 
and in quantifying structures of the Galactic H{\sc i} emission \citep{kalberla2023}. 
They have also been used to study the morphology of turbulent fluid 
flows \citep{calzavarini2008}, magnetic field structures in the kinematic \citep{wilkin2007} 
and nonlinear stages \citep{Seta+20} of fluctuation dynamos, and even those of 
reconnecting fields \citep{DAB24} in magnetically dominated decaying turbulence.
Recently, they have also been used to infer the shapes of substructures in shocks 
of radio relics \citep{WBGR23}.

In \citet{SBS21} and \citet{BS21}, we studied in detail the statistical properties 
of polarized synchrotron emission arising from fluctuation dynamo action in the 
context of the intracluster medium (ICM), which is both Faraday rotating and 
synchrotron emitting. In this work, we address for the first time certain key questions 
that provide crucial insights into the morphology of these polarized structures. With 
the aid of synthetic maps of the fractional polarization ($\pf$), we explore how the 
number of 2D structures varies with $\pf$ and the effects of frequency-dependent 
Faraday depolarization on the number and filamentarity of these structures when 
the size of the emission regions are larger compared to the turbulent driving scales 
($\lf$). Additionally, we also explore the effect of emission regions on scales comparable 
to or smaller than $\lf$, and explore how the morphological features are related to $\lf$. 
As we show in this paper, our analysis reveals that Minkowski functionals are effective 
probes to study the properties of polarized synchrotron emission, and the number of 
such structures can indeed be used to infer the scale of turbulent motions.
 
\section{Data and methodology}
\label{s:data}

\subsection{Simulations and synthetic observations}

We use data from three non ideal MHD simulations of fluctuation dynamos reported 
in \citet{BS21} where turbulence is driven solenoidally (i.e., $\nabla\cdot\FF = 0$, $\FF$ 
is the forcing term) over a range of wave numbers --- (i) $1 \leq |\kk|L/2\uppi \leq 3$, 
(ii) $4 \leq |\kk|L/2\uppi \leq 6$, and (iii) $7 \leq |\kk|L/2\uppi \leq 9$, as a 
stochastic Ornstein--Uhlenbeck process with a finite time correlation. Thus, the 
average forcing wavenumbers are $k_{\rm f}L/2\uppi = 2$, $5$, and $8$, where 
$L$ is the length of the box. These $\kf$ correspond to turbulent driving scales 
$\lf = 2\uppi/\kf = L/2$, $L/5$, and $L/8$, respectively. We refer the readers to 
\citet{BS21,SBS21} and \citet{SS24} for more details on the basic equations, 
numerical setup, and initial conditions. In summary, we solve the full set of 3D 
magnetohydrodynamic equations in dimensionless units on a uniform grid consisting 
of $512^{3}$ grid points in a periodic box of unit length with an isothermal equation 
of state using the {\texttt{FLASH}} code\footnote{\url{https://flash.rochester.edu/site/flashcode/}}
\citep{Fry+00,Benzi+08, EP88} (version 4.2). 
The amplitude of the driving results in subsonic turbulence with rms Mach number 
$\mathcal{M}\approx 0.2$. The eddy turnover time at the driving scale is 
$\ted = \lf/\urms$, where $\urms$ is the rms value of the turbulent velocity in the
steady state. For given values of the physical viscosity ($\zeta$)\footnote{Note that 
performing nonideal MHD simulations is paramount for fluctuation dynamos as this 
mechanism essentially involves processes close to resistive scales. Hence, we use 
physical values of viscosity ($\nu$) and resistivity ($\eta$) in the momentum and 
magnetic induction equations, respectively. This is in contrast to ideal MHD simulations 
where both viscous and resistive dissipation are controlled by numerical diffusion alone.}, 
we obtain fluid Reynolds number $\Rey = \urms\lf/\zeta = 1080,1450$ and $1425$ in our 
simulations with $\lf = L/2$, $L/5$, and $L/8$, respectively. These 
are large enough to ensure that the flows are turbulent\footnote{
Following \citet{MSL20}, the Reynolds number corresponding to numerical diffusion 
is $\approx 2N_{\rm grid}^{4/3}\approx 8000$ for subsonic simulations having 
$N_{\rm grid} = 512$ grid points. This implies that explicit diffusion is significantly larger 
than numerical diffusion in all our runs.}.

Each of the aforementioned simulations was run for several eddy turnover 
times so as to capture the kinematic, intermediate, and saturated phases of dynamo 
evolution. The existence of an intermediate phase has been reported in 
earlier studies \citep[e.g.,][]{Scheko+04PRL,Cho+09,Beres12,SS24}. This phase is 
expected to commence when the magnetic energy density becomes comparable to 
the kinetic energy density on the smallest supercritical scales and end when the 
above energy densities become comparable on scales near the outer scale of 
turbulence \citep{SS21}. To address the objectives of our study, we solely focus on a 
number of independent realizations in the saturated phase of the dynamo. This 
phase is captured over $15, 16$, and $20\,\ted$ in simulations with $\lf = L/2$, $L/5$, 
and $L/8$, respectively. The morphological analysis is then performed by utilizing
multiple 3D snapshots of the gas density ($\rho$) and the three components of the 
magnetic field ($B_{x}, B_{y}, B_{z}$) for computing 2D $512\times512$\,pixel maps 
of the total synchrotron ($I$), and the polarized synchrotron intensity ($\PI$) in the 
plane of the sky using the {\texttt{COSMIC}} package \citep{basu19b}. Physical 
unit conversion, the choice of the cosmic ray energy spectrum, and normalization 
for the synchrotron intensity are identical to those described in \citet{SBS21} and 
\citet{BS21}. We have chosen the $x$- and $y$-axes to be in the plane so that the 
magnetic field component in the plane of the sky $B_{\perp} = ({B_{x}^{2} + B_{y}^{2}})^{1/2}$ 
contributes to the total and polarized synchrotron emission, and the component 
parallel to the LOS, $B_\|=B_z$, contributes to Faraday rotation. In this work, we 
will focus on the analysis of 2D maps of fractional polarization ($\pf$) at a frequency 
$\nu$, defined as $p_{\rm f,\nu} = PI_\nu/I_\nu$, because it is largely independent 
of our fiducial choice of the normalization of $I$ and the assumed shape of the 
frequency spectrum. Furthermore, we use synthetic maps at three representative 
frequencies $\nu = 6, 1$, and $0.5\ghz$, where the effects of frequency-dependent 
Faraday depolarization are negligible, moderate, and strong, respectively.

We would like to emphasize that the value of the Faraday rotation measure (RM) and
its dispersion ($\sigma_{\rm RM}$) depends on the choice of the physical size of the 
simulation domain $L$ and the free-electron number density ($n_{\rm e}$). The synthetic 
maps of RM have $\sigma_{\rm RM}$ in the range 90--120\,rad\,m$^{-2}$ \citep[][]{BS21}, 
and therefore, the frequency dependence of Faraday depolarization would differ for other 
choices of $L$ and $n_{\rm e}$. Although the qualitative results would remain unaffected, 
the results at $0.5$ and $1\ghz$ should be considered as representative frequencies 
where Faraday depolarization is high and moderate, respectively. However, all our 
conclusions are unaffected at $\nu \gtrsim3\ghz$.

\subsection{Computing Minkowski functionals}

To quantify the morphology of 2D structures in the maps of $\pf$, we computed 2D 
{\it Minkowski functionals} -- the area ($A$) and perimeter ($P$), by developing a new, 
computationally efficient algorithm implemented via a Python-based software, 
\texttt{perimetrics} \citep{perimetricsv1_0_0}\footnote{\url{https://zenodo.org/records/11118211}}.
This is based on the {\it Marching Squares} algorithm of \citet{mantz08}. 
The workflow and tests of the algorithm are described in the Appendix.
Given $A$ and $P$, the shape of a structure is quantified in 
terms of a single dimensionless number, the "filamentarity" ($F$), defined as,
\EQ
F=\frac{P^2 - 4 \,\uppi\, A}{P^2 + 4\, \uppi\, A}.   
\label{eq:filament_defn}
\EN 
The value of $F$ is restricted between 0 and 1, such that $F = 0$ for a circle, and 
$F = 1$ for a 1D line that can be either straight or curved. This also implies that 
$F$ is not an additive quantity. Further, it is important to note that, by definition, 
$F$ is sensitive to the shape of the structures but not to their size. In the following, 
we study $F$ of 2D maps of $\pf$ for structures above various threshold values 
of $\pf$.

\section{Results}
\label{s:results}

Utilizing Minkowski functionals, we first compare the morphology of polarized 
structures generated at varying turbulence driving scales ($\lf$) and then study 
the effects of Faraday depolarization on these structures.

\subsection{Number of connected emitting components}
\label{subsec:num_emit}

%%%%%%%%%%%%%%%%%%%%%%%%%%%%%%%%%%%%%%%%%%%%%%%%%%%%%%%%%%%%%%%%%%%%
\begin{figure}[t!]
\begin{centering}
\includegraphics[width=\columnwidth]{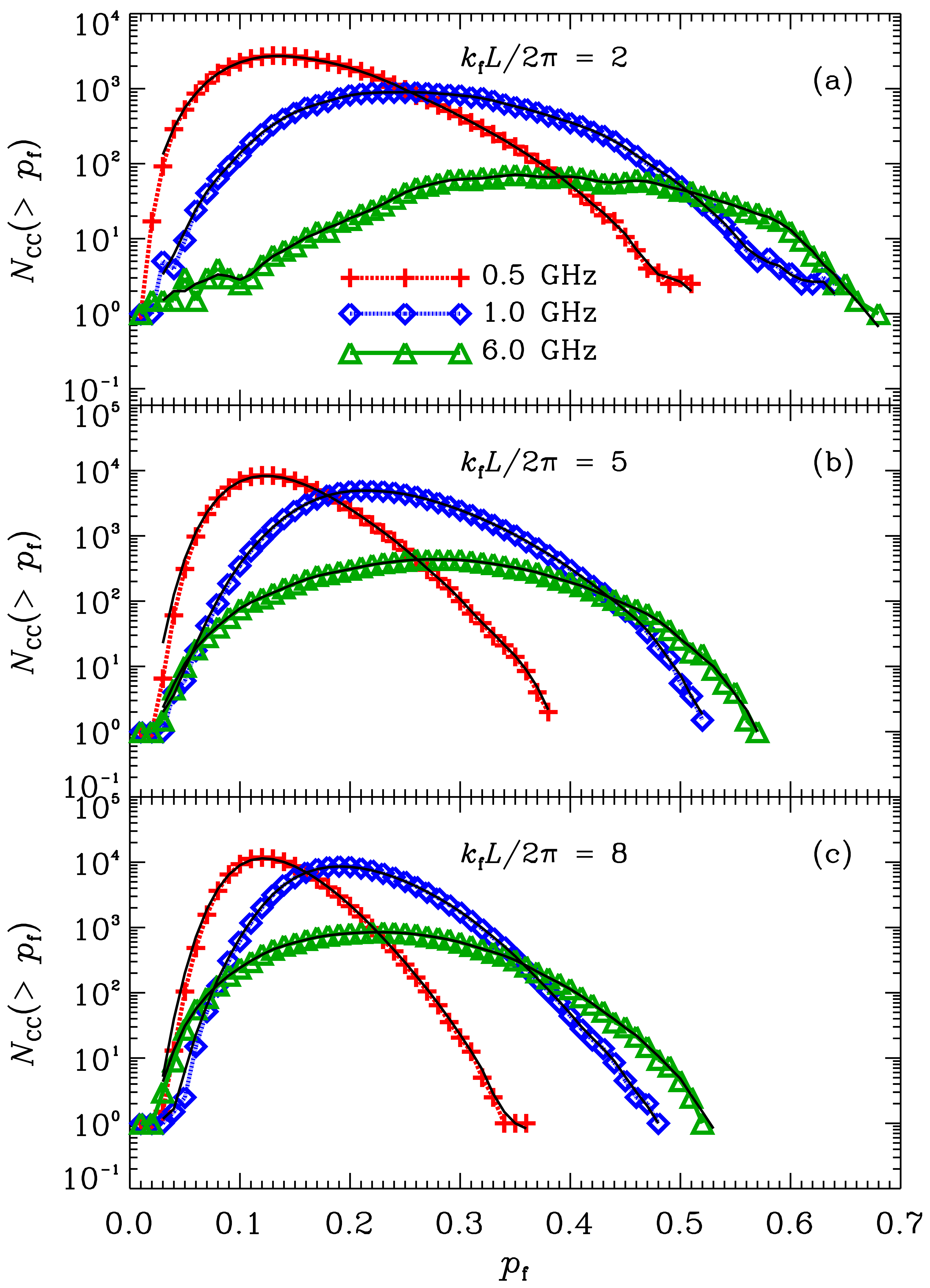}
\end{centering}
\caption{Variation of $\ncc$ with the threshold $\pf$ at $\nu = 0.5\ghz$ 
(red, "plus"), $1.0\ghz$ (blue, "diamond"), and for $\nu = 6.0\ghz$ 
(green, "triangle") for simulations with different driving scales: $\kf L/2\pi= 2$ 
(panel, "(a)"), $5$ (panel "(b)"), and $8$ (panel "(c)"). The $N_{\rm cc}$ 
values for each curve were computed over a range of snapshots in the 
saturated phase of the dynamo in each run. The black solid curves are 
those obtained by smoothing the data.}
\label{fig:number_conn}
\end{figure}
%%%%%%%%%%%%%%%%%%%%%%%%%%%%%%%%%%%%%%%%%%%%%%%%%%%%%%%%%%%%%%%%%%%%%%%%

Using 8-connectivity \citep{GW02}, we determine the {\it connected components}
(see Appendix~\ref{sec:algo}) above a threshold value of $\pf$ from the synthetic 
maps. This allows us to filter continuous polarized structures above $\pf$, and 
compute $A$ and $P$ for each of them. In Figure~\ref{fig:number_conn}, 
we show the number of connected components ($\ncc$) above a threshold 
$\pf$, $\ncc(>\pf)$, as a function of $\pf$ at $\nu = 0.5$, $1$, and $6\ghz$ as 
the different symbols, and for the three $\lf$s as the different 
panels. For independence, the data points shown in Figure~\ref{fig:number_conn} 
are averaged for two snapshots that are separated by more than $\ted$.
Firstly, we find that, in all cases, $\ncc(>\pf)$ 
peaks for a certain value of $\pf$, such that the peak is significantly broader at 
higher frequencies compared to that at lower frequencies. We define $\ppeak$ 
as the $\pf$ where $\ncc(>\pf)$ is maximum.\footnote{Since the location of 
$\ppeak$ could be affected by stochastic variations in $\ncc$, seen as small 
fluctuations at higher frequencies in Figure~\ref{fig:number_conn}, we determine 
$\ppeak$ by fitting the peak of the top-hat smoothed curves with a parabola.} 
Above $\ppeak$, $\ncc$ decreases rapidly, especially below $\approx1\ghz$. 
Thus, $\ppeak$ represents the transition value at which the breaking up of 
larger structures is balanced by the shrinking and disappearance of smaller 
structures. Secondly, irrespective of $\kf$, the highest number of connected 
structures, measured in terms of $\ncc$, always occurs at low frequencies, 
here at $0.5\ghz$, for a threshold $\pf \leq 0.15$, close to $\pmean = 0.11$ 
determined from the $\pf$ map at $0.5\ghz$. This is a direct consequence of 
Faraday depolarization, which gives rise to a plethora of structures at low $\pf$. 
Thirdly, at a given frequency, $\ppeak$ decreases with increasing $\kf$, similar 
to what was seen for $\pmean$ in \citet{BS21}.

\begin{figure}[t!]
\begin{centering}
\includegraphics[width=\columnwidth]{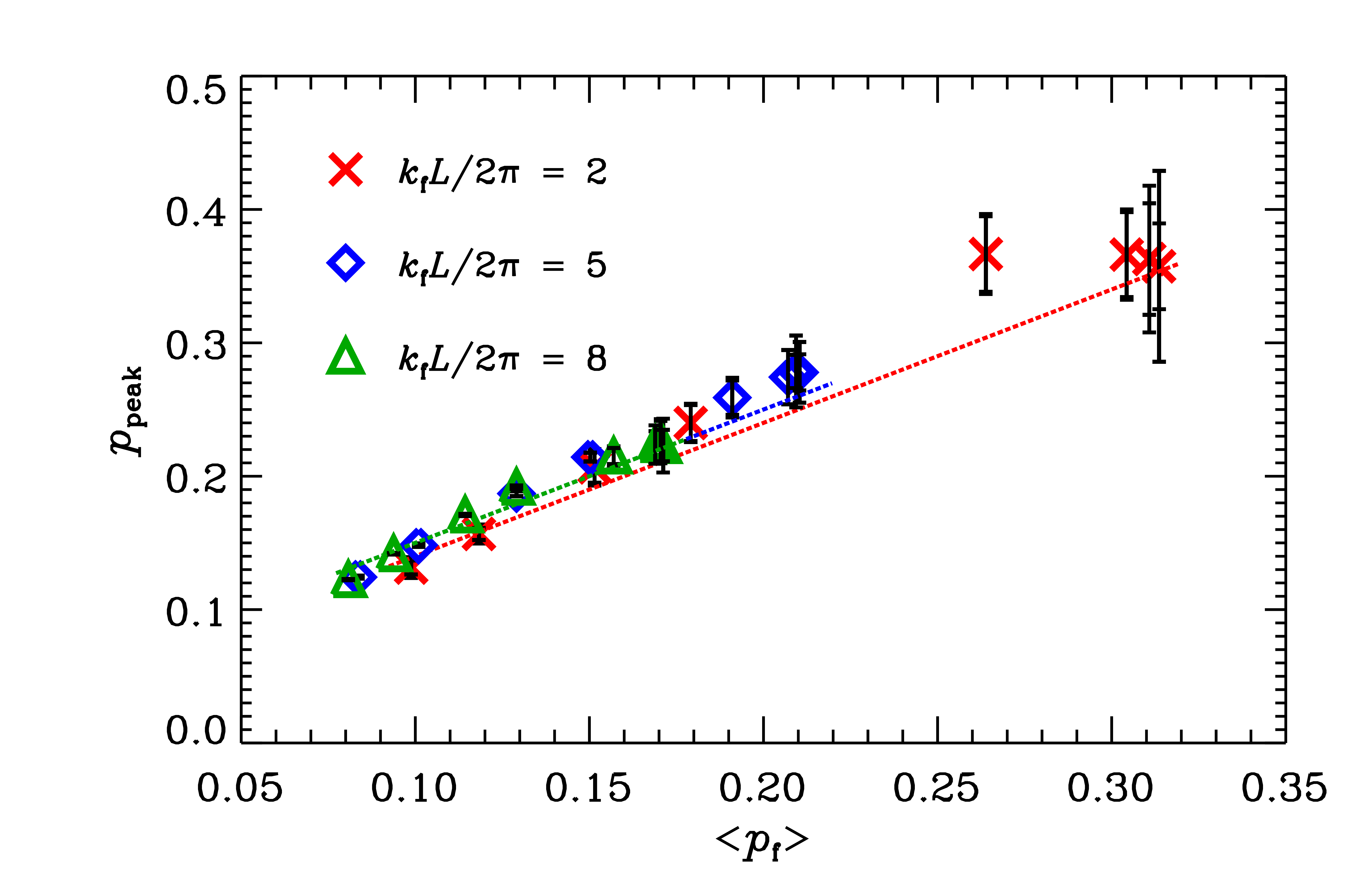} 
\end{centering}
\caption{Relation between $\pmean$ and median values of $\ppeak$ 
corresponding to $\nu = 0.5, 0.6, 0.8, 1.0, 2.0, 4.0, 5.0$, and $6.0\ghz$ 
(data points from left to right) along with $2\sigma$ error bars.
Red crosses: $\kf L/2\pi=2$, blue diamonds: $\kf L/2\pi=5$, and 
green triangles: $\kf L/2\pi=8$. The colored dotted lines represent the 
1:1 linear fit to the data.}
\label{fig:pf_mean_peak}
\end{figure}

The physical connection between $\ncc$ and $\pf$ can be understood from 
Figure~\ref{fig:pf_mean_peak}, which shows the variation of $\ppeak$ with 
$\pmean$ directly determined from the synthetic maps at different frequencies
for all three driving scales. Remarkably, we find that $\ppeak$ and $\pmean$ 
closely follow a 1:1 relation (dotted lines in Figure~\ref{fig:pf_mean_peak}), 
indicating $\ppeak$ to be a direct measure of $\pmean$ within $\approx10\%$.
However, for $\nu \gtrsim3\ghz$, both saturate to the value expected for the 
intrinsic value of $\bra{\pf}$ \citep[][]{SBS21, BS21}.

%%%%%%%%%%%%%%%%%%%%%%%%%%%%%%%%%%%%%%%%%%%%%%%%%%%%%%%%%%%%%%%%%%%%%%%%%%
%%%%%%%%%%%%%%%%%%%%%%%%%%%%%%%%%%%%%%%%%%%%%%%%%%%%%%%%%%%%%%%%%%%%%%%%%%%

%%%%%%%%%%%%%%%%%%%%%%%%%%%%%%%%%%%%%%%%%%%%%%%%%%%%%%%%%%%%%%%%%%%%%%%%%%%%%%%%
\begin{figure*}
\hspace{-45pt}
\begin{centering}
\begin{tabular}{c c c}
\Large{$\kf L/2\uppi=2$} & \Large{$\kf L/2\uppi=5$} & \Large{$\kf L/2\uppi=8$} \\
&  \\
\multicolumn{3}{c}{\Large{6~GHz}}  \\
{\mbox{\includegraphics[width=6.0cm]{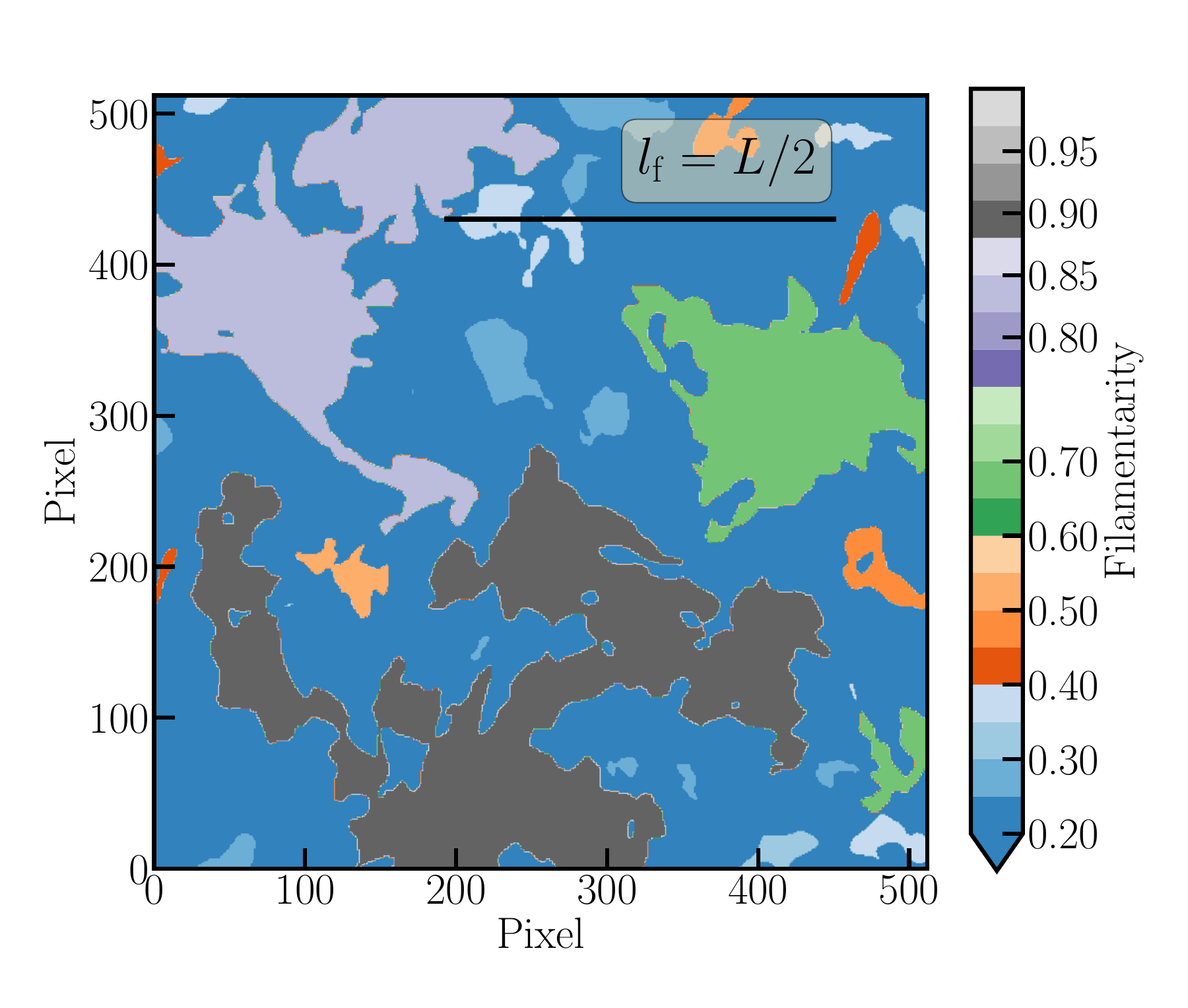}}} & 
{\mbox{\includegraphics[width=6.0cm]{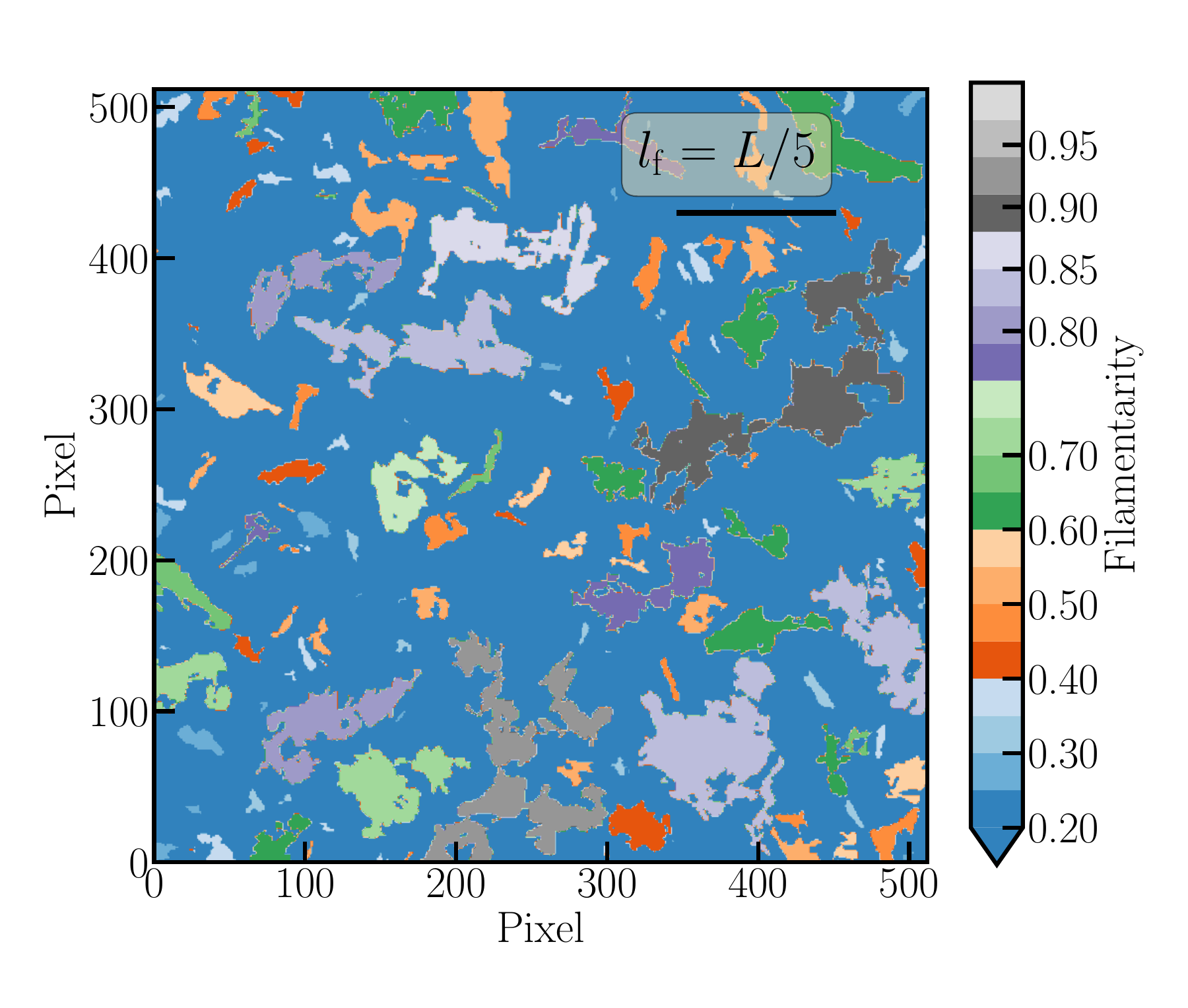}}} &
{\mbox{\includegraphics[width=6.0cm]{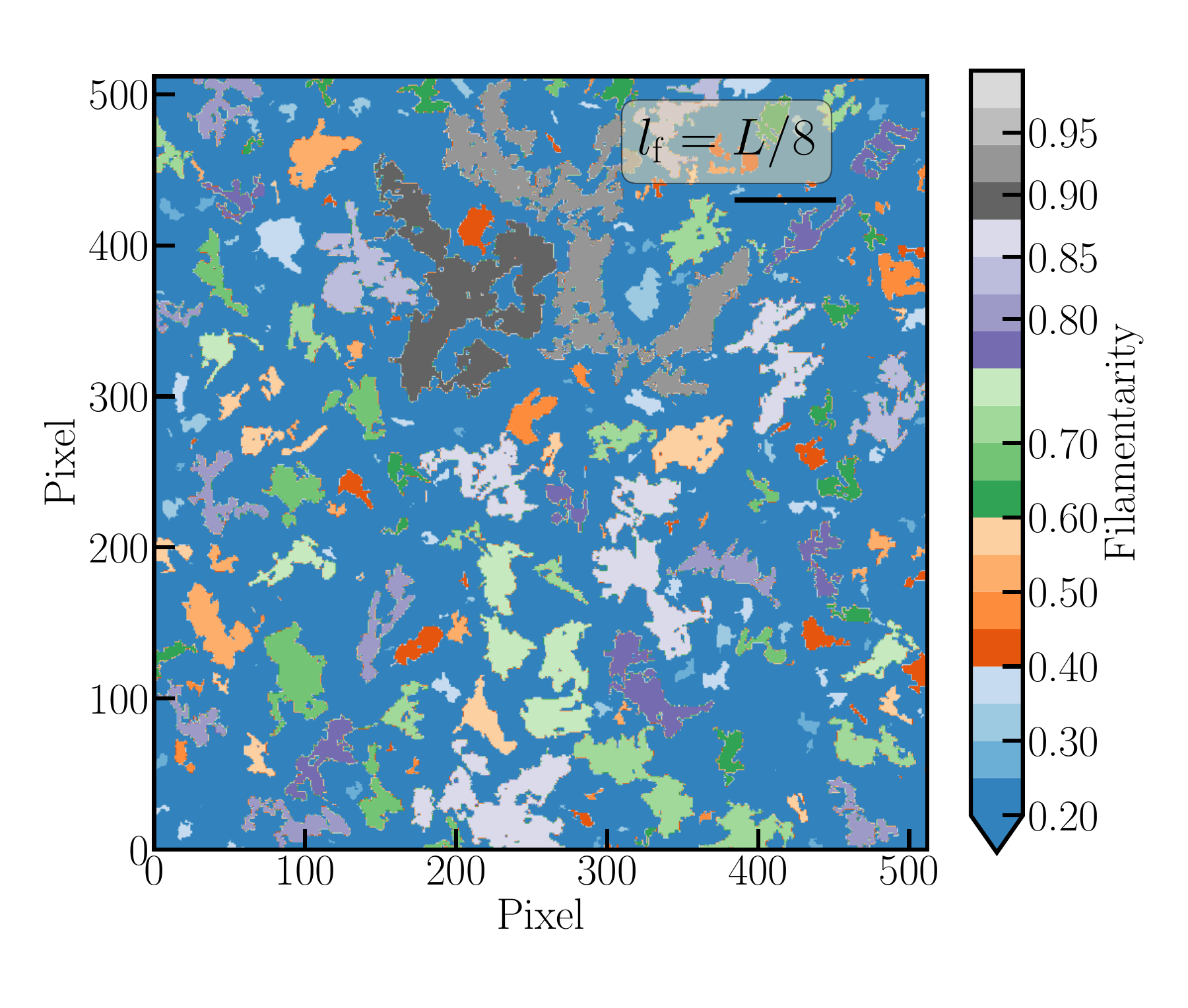}}}\\
\multicolumn{3}{c}{\Large{0.5~GHz}}  \\
{\mbox{\includegraphics[width=6.0cm]{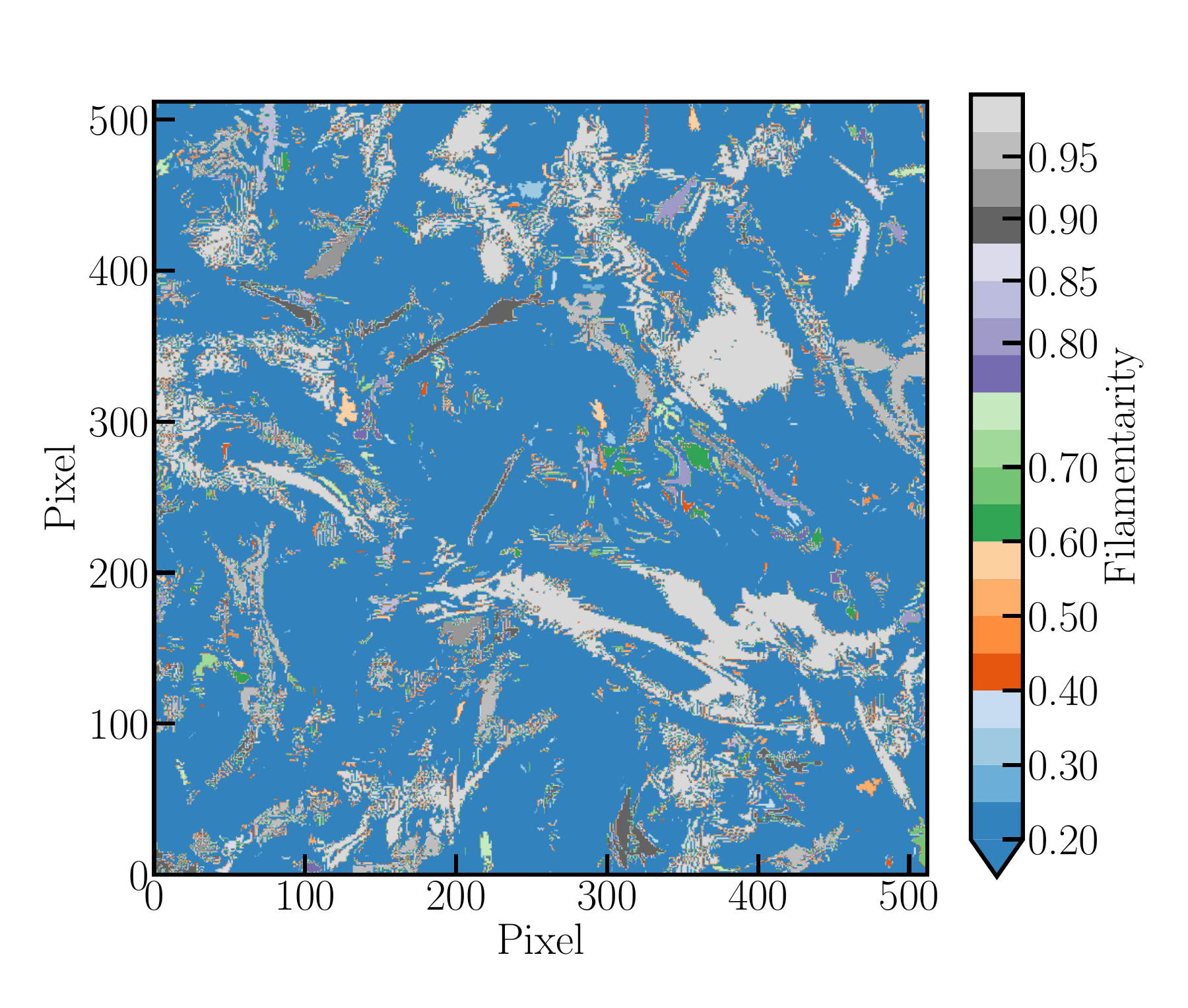}}} &
{\mbox{\includegraphics[width=6.0cm]{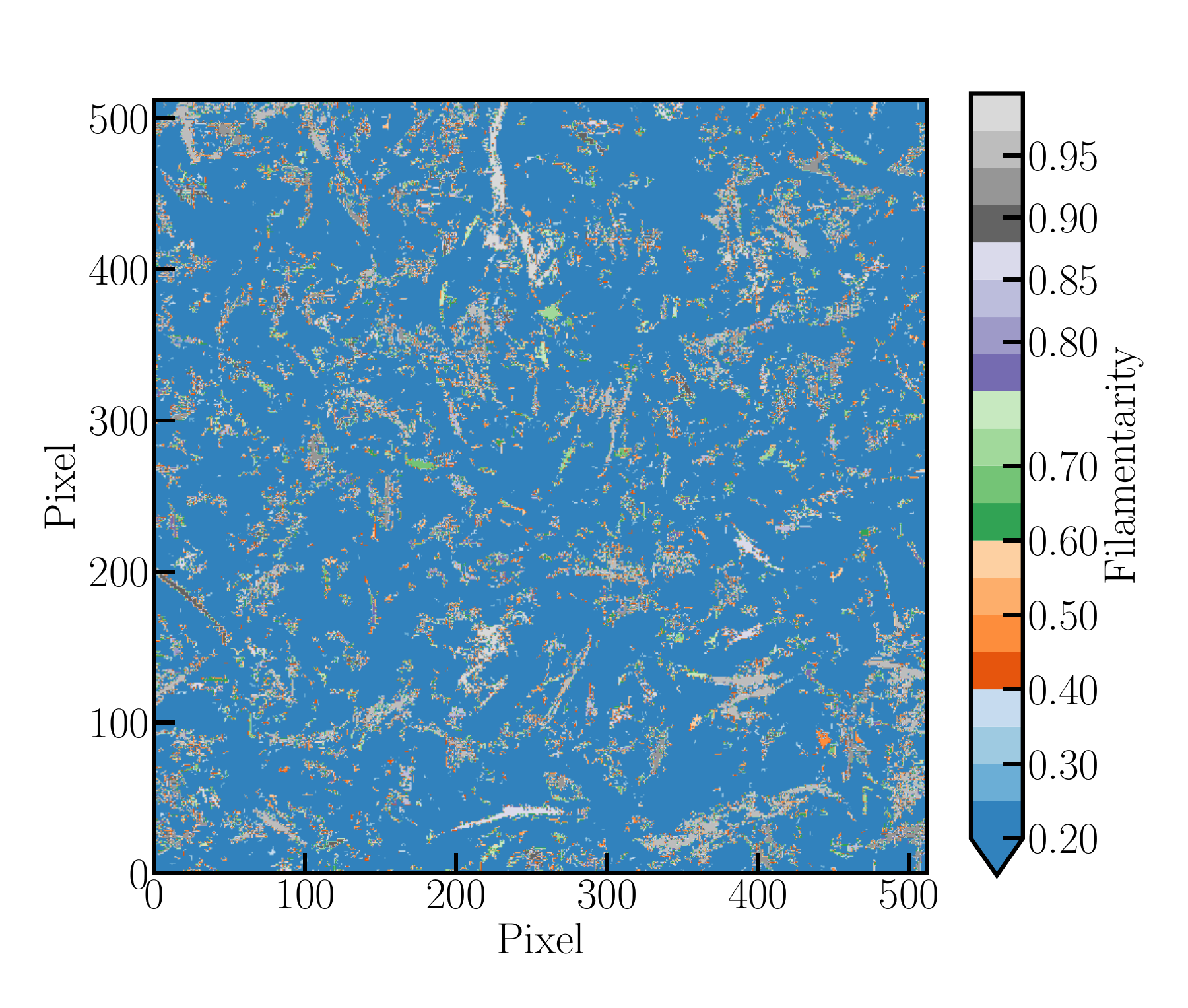}}} &
{\mbox{\includegraphics[width=6.0cm]{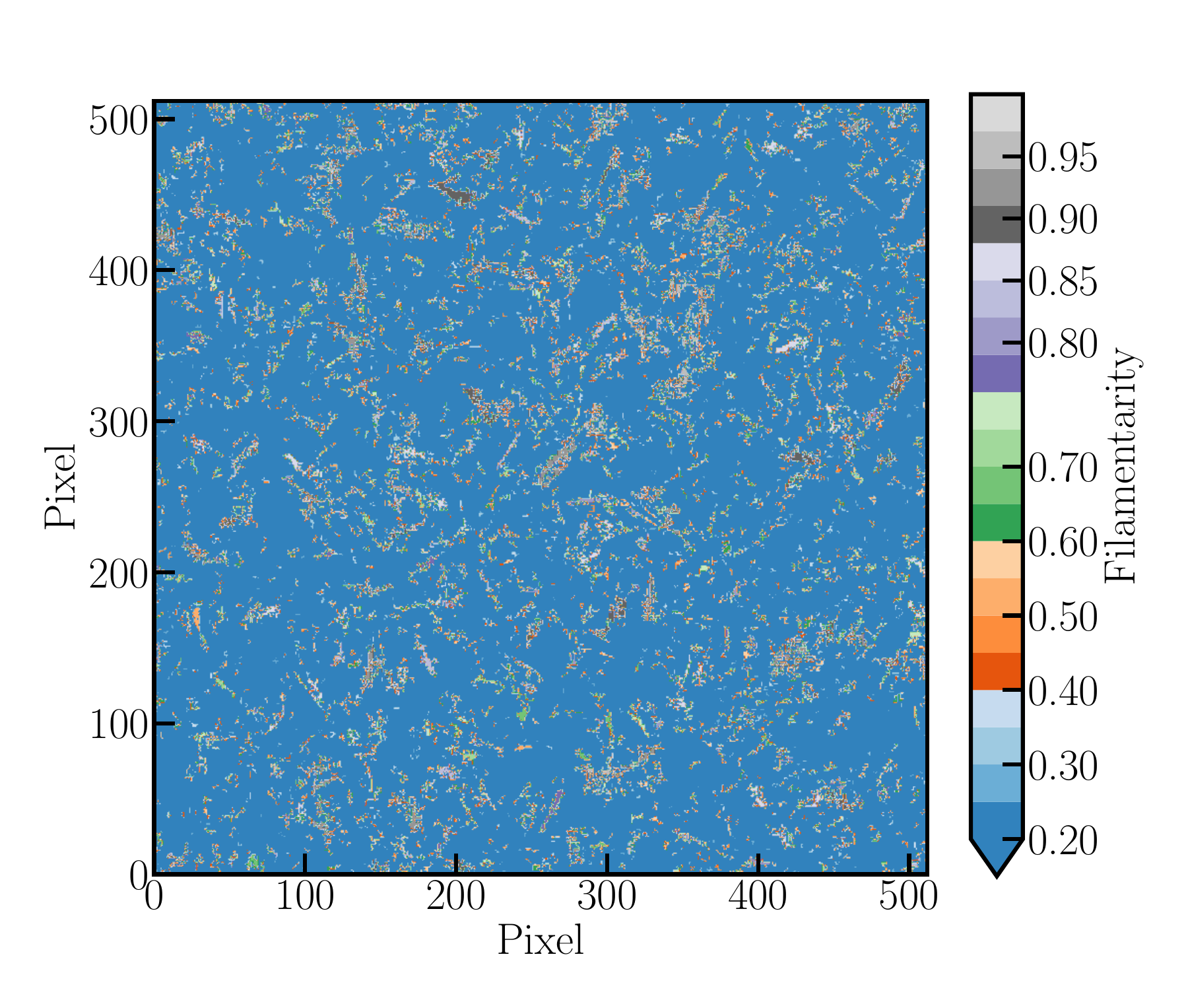}}} \\
{\mbox{\includegraphics[width=5cm]{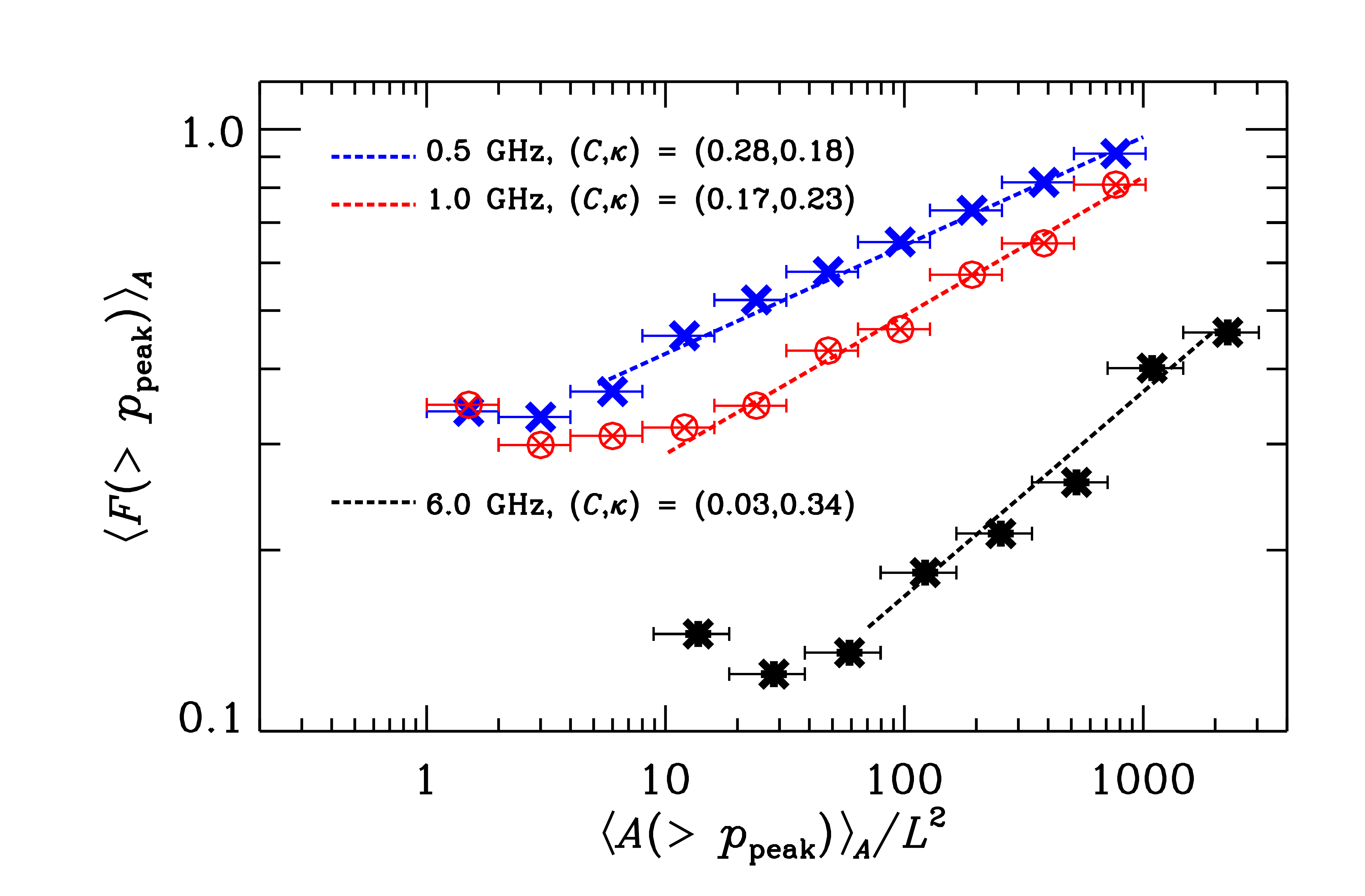}}}&
{\mbox{\includegraphics[width=5cm]{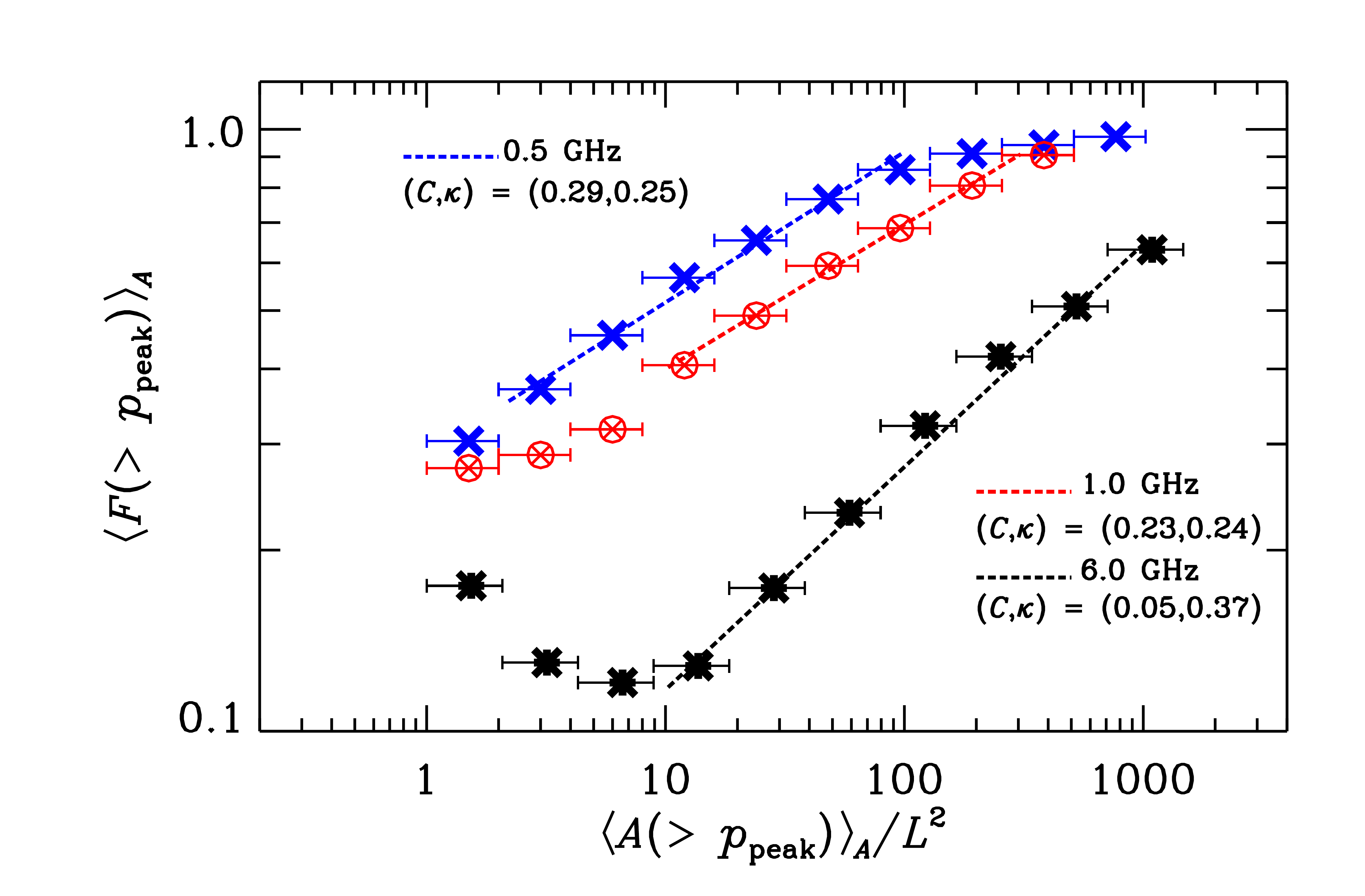}}}&
{\mbox{\includegraphics[width=5cm]{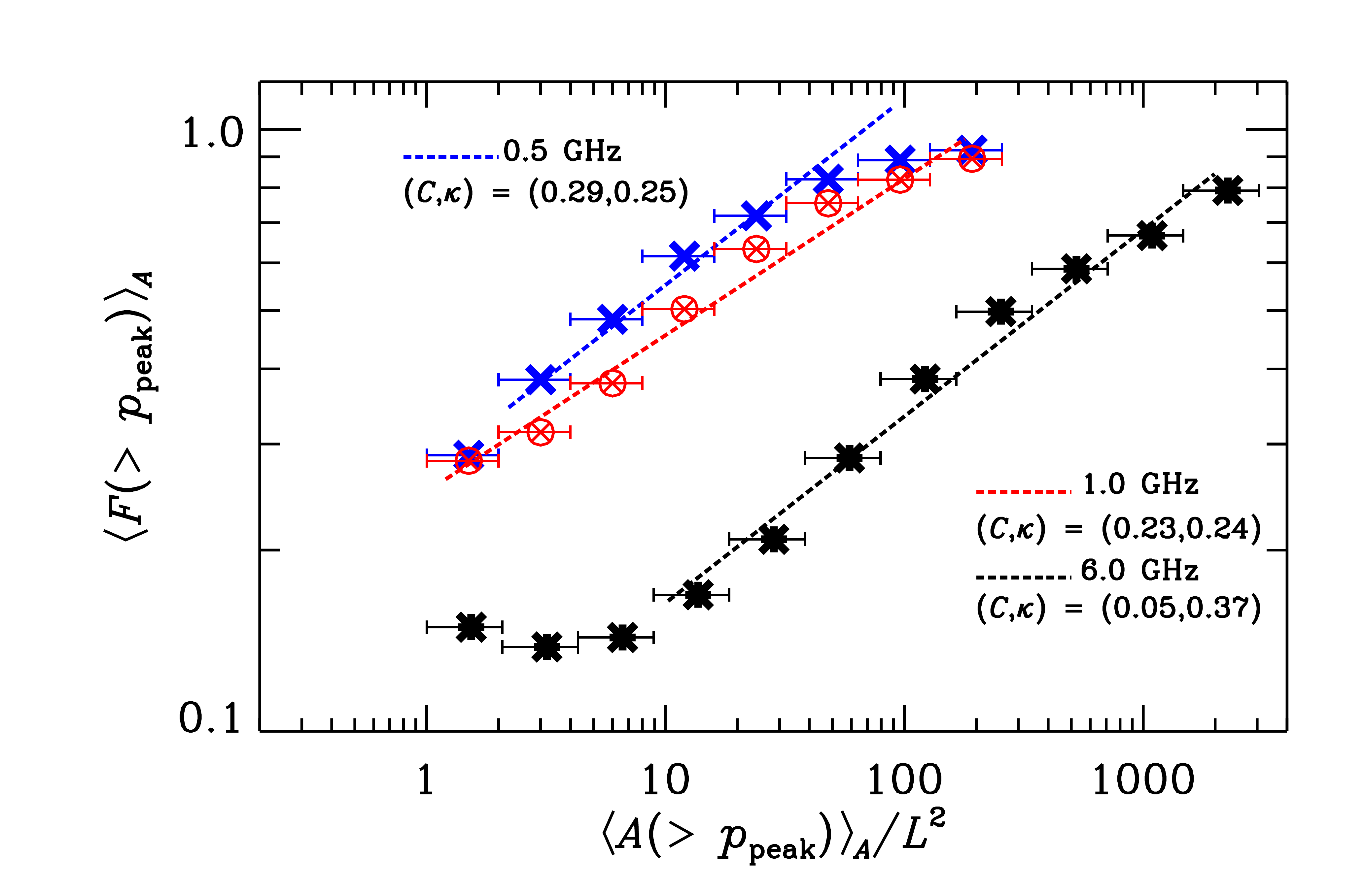}}}\\
\end{tabular}
\end{centering}
\caption{Top and middle rows: Maps of filamentarity above the fractional polarization 
threshold of $\ppeak$ (see Section~\ref{subsec:num_emit} for details). The top and 
middle rows are for $\nu = 6$ and $0.5\ghz$. Bottom row: Variation of average 
filamentarity in each area bin $\bra{F(>\ppeak)}_A$ with average area $\bra{A(>\ppeak)}_A$ 
binned logarithmically. The dashed lines are linear fits to these curves, for which the 
$x$-coordinates are taken to be the bin centers. The left, middle, and right columns 
are for turbulence driven with $\kf L/2\uppi = 2, 5$, and $8$, respectively. 
}
\label{fig:fil_image_map}
\end{figure*}
%%%%%%%%%%%%%%%%%%%%%%%%%%%%%%%%%%%%%%%%%%%%%%%%%%%%%%%%%%%%%%%%%%%%%%%%%%%%%%%%%

\subsection{Filamentarity and sizes of emitting components}
\label{subsec:filamen_emit}

Here we present the results of the filamentarity of the polarized structures and 
its dependence on different threshold values of $\pf$ and area $A$. Note that it 
is difficult to measure $A$ and $P$ accurately for structures on the scale of the 
grid points, as they are sensitive to the details of the interpolation and contouring 
schemes. Hence, we only considered structures having $A > 1$\,pixel$^{2}$ 
throughout this study. Figure~\ref{fig:fil_image_map} shows the map of 
$F(>\ppeak)$ at 6\,GHz (top row) and 0.5\,GHz (middle row) for the three turbulent 
driving scales. It is apparent that polarized structures with a wide range in $F$ 
occur in all three cases.For a given $\kf$, Figure~\ref{fig:fil_image_map} shows 
that larger structures on scales comparable to $\lf$ with lower $F$ at $\nu = 6\ghz$ 
fragment into smaller structures with higher $F$ at $\nu = 0.5\ghz$ due to stronger 
Faraday depolarization at $\nu\lesssim1\ghz$. This is seen for all choices of the 
threshold value of $\pf$. Notably, from these synthetic data, we find the mean 
filamentarity of all the connected components above a $\pf$, $\langle F(>\pf)\rangle$,
to remain mostly constant as a function of $\pf$. In fact,  $\langle F(>\pf)\rangle$ is 
consistently larger with $\langle F\rangle >0.3$ than that at higher frequencies 
($\gtrsim3\ghz$), where $\langle F\rangle$ lies in the range $0.15\textrm{--}0.22$ 
for all the driving scales, and for all choice of threshold $\pf\gtrsim0.05$.\footnote{As seen in 
Figure~\ref{fig:number_conn}, the number of $\ncc$ at $\pf \lesssim0.05$ is low 
for $\kf L/2\uppi = 5$ and $8$, and therefore, the variation of $\bra{F}$ in this 
regime is subject to large statistical fluctuations.} 

To see the size dependence of the polarized structures with $F$, in the bottom row 
of Figure~\ref{fig:fil_image_map} we show the variation of $F$ averaged within 
logarithmic bins of $A$, $\bra{F(>\ppeak)}_A$, as a function of the mean area 
$\bra{A(>\ppeak)}_A$. Here, the subscript "$A$" represents averaging of the quantities 
for all connected components whose area, in pixels, lie in the range $\Delta A^{n\,K}$ 
and $\Delta A^{(n+1)\,K}$, where $n\in \mathbb{Z}^+$, $K=0.5$, and 
$\Delta A = 4$\,pixels$^2$ for $0.5$ and $1\ghz$, and $\Delta A=4.3$\,pixels$^2$ 
for $6\ghz$. To reduce statistical fluctuations of $\bra{F(>\ppeak)}_A$, especially
 in the largest area bins that typically have few connected structures, the data points 
 in Figure~\ref{fig:fil_image_map} (bottom row) are further averaged for two independent 
snapshots that are separated by $>t_{\rm ed}$. Clearly, as a consequence of increased 
Faraday depolarization at lower frequencies, $\bra{F(>\ppeak)}_A$ is larger for all the 
driving scales for structures spanning a large range in the area. Furthermore, for 
structures that have $A\gtrsim10$\,pixels$^2$, $\bra{F(>\ppeak)}_{A}$ varies with 
$\bra{A(>\ppeak)}_A$ roughly as a power law, $\bra{F(>\ppeak)}_A = C\,\bra{A(>\ppeak)}_A^\kappa$. 
The values of the normalization $C$ and power-law index $\kappa$ are shown in 
the bottom row of Figure~\ref{fig:fil_image_map}. We find $\kappa$ to lie in the range 
0.18 and 0.35, implying large-scale polarized structures to be more filamentary on 
average. Finally, for $\nu \gtrsim3\ghz$, where the polarized emission is similar to 
the intrinsic emission \citep{BS21}, $\kappa$ remains roughly the same, 
$\kappa = 0.31\textrm{--}0.37$, for the three $\kf$s. However, the normalization $C$ 
progressively decreases with increasing $\kf$ by more than a factor of 2. This 
indicates that, for turbulence driven on small scales, the volume-filling polarized 
structures \citep{BS21} are significantly more filamentary. On the other hand, at 
$\nu\lesssim1\ghz$, the exponent shows a steady increase from 
$\kappa=0.18$ to $0.31$, for $\kf$ increasing from $2$ to $8$. Interestingly, we 
find indication that, at frequencies where Faraday depolarization is high, e.g., 
at $0.5\ghz$ in our case, the area of the largest structures is more than 10 
times smaller than those at $\gtrsim3\ghz$ for turbulence driven on small scales.

\subsection{Dependence on the projected size in the plane of the sky}

In Sections~\ref{subsec:num_emit} and \ref{subsec:filamen_emit}, we have 
performed the morphological analysis for the entire simulated domain, i.e., 
for the case when polarized emission is detectable over the entire 
$512\times512$\,pixels$^{2}$ 2D maps. Here, we study the impact on our 
results for subregions of various sizes ($\ell$), especially on the determination 
of $\ppeak$ at 6\,GHz. This is equivalent to cases where synchrotron emission is 
detectable only from a fraction of the emitting area, e.g., because of insufficient 
telescope sensitivity. This allows us to estimate the scales of $\ell$ at which the 
linear relationship between $\ppeak$ and $\pmean$ seen in Figure~\ref{fig:pf_mean_peak} 
becomes unreliable. To this end, we sampled multiple regions of size $\ell\times \ell$ 
in the 2D maps of $\pf$ for $\ell$ ranging from 25 to 512\,pixels. These correspond 
to $\ell/\lf$ ranging from $1/8$ to $8$ for the different $\kf$. For example, 
$\ell = 100$\,pixels corresponds to an area of $\ell/\lf\times\ell/\lf \approx 2/5\times2/5$ 
for $\kf\,L/2\uppi = 2$, $\approx1\times1$ for $\kf\,L/2\uppi = 5$, and $\approx8/5\times8/5$ 
for $\kf\,L/2\uppi = 8$. We chose three different locations for the subregions for each 
$\kf\,L/2\uppi$, and for the case of $\ell/\lf=2$ for $\kf\,L/2\uppi = 2$, we used two 
different snapshots in the saturated stages of the turbulent dynamo.

%%%%%%%%%%%%%%%%%%%%%%%%%%%%%%%%%%%%%%%%%%%%%%%%%%%%%%%%%%%%%%%%%%%
\begin{figure}
\hspace{-30pt}
\begin{centering}
\begin{tabular}{c}
{\mbox{\includegraphics[height=6.3cm, trim=0cm 0cm 0cm 0cm, clip]{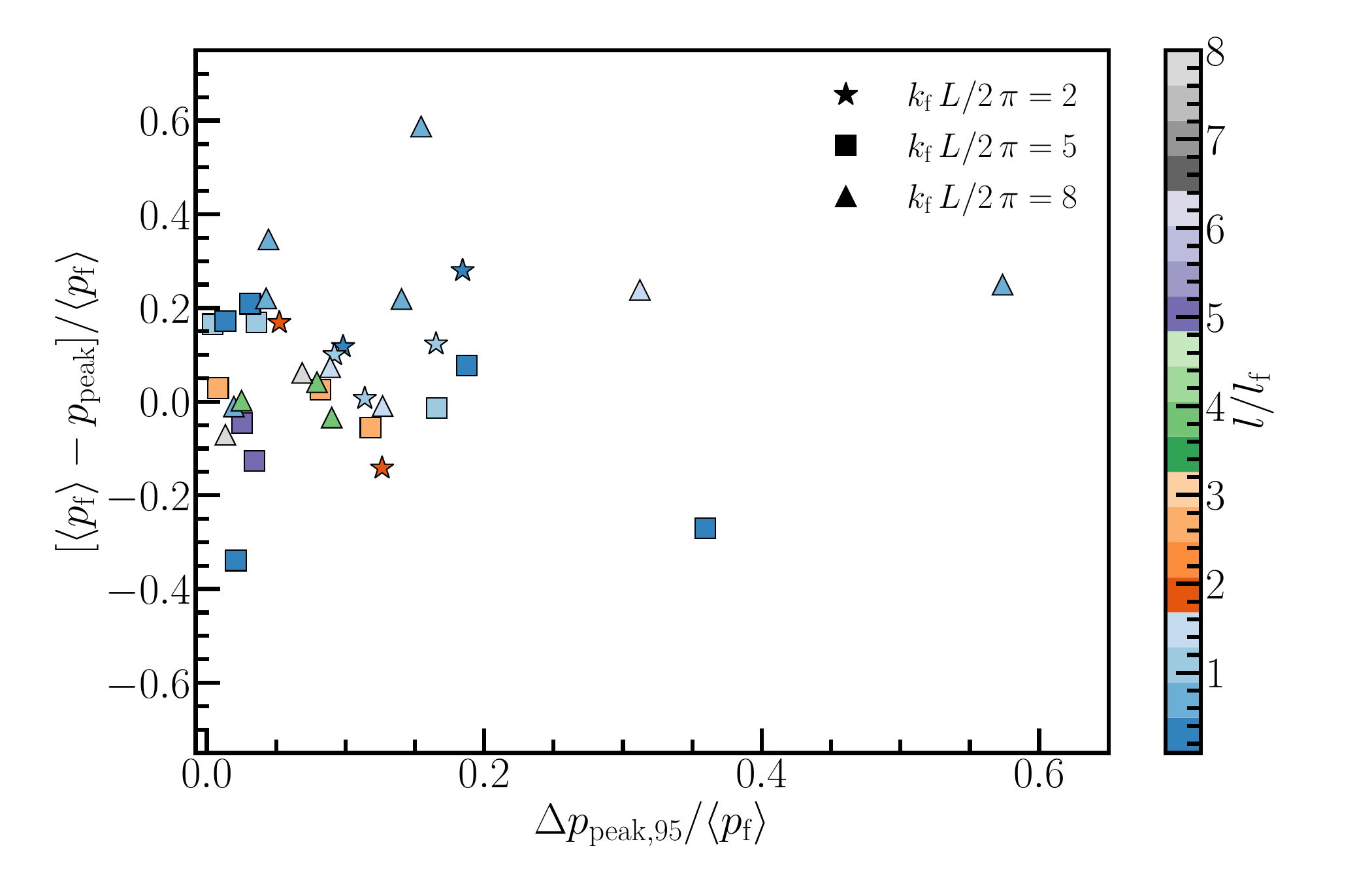}}}\\
\end{tabular}
\end{centering}
\caption{Variation of the relative error on the estimation of $\ppeak$, $[\bra{\pf} - \ppeak]/\bra{\pf}$, 
as a function of relative 95 percentile range in $\ppeak$, $\Delta p_{\rm peak,95}/\bra{\pf}$, 
determined from Monte Carlo sampling at 6\,GHz. The star, square, and triangle data points 
represent $\kf L/2\uppi = 2,5$, and 8, respectively. Here, $\pmean$ is determined for the entire 
emitting region. The points are colorized based on the subgrid size expressed in $\ell/\lf$.}
\label{fig:ppeakerror}
\end{figure}
%%%%%%%%%%%%%%%%%%%%%%%%%%%%%%%%%%%%%%%%%%%%%%%%%%%%%%%%%%%%%%%%%%%%%

In Figure~\ref{fig:ppeakerror}, we show the variation of the relative error on the
estimated $\ppeak$ for different $\ell/\lf$ as a function of the 95\,percentile range, 
$\Delta\,p_{\rm peak,95}$, of the $\ppeak$ determined from Monte Carlo samples.\footnote{We 
generated 100 samples from the $\ncc(>\pf)$ \textit{versus} $\pf$ plot for each subgrid region
(similar to Figure~\ref{fig:number_conn}) by drawing each $\ncc$ from a Poisson 
distribution and estimated $\ppeak$ by fitting each random sample in the same 
way as for Figure~\ref{fig:number_conn}.} Both these quantities are compared 
relative to $\pmean$ over the entire $512\times512$\,pixel$^2$ 2D map. In 
Figure~\ref{fig:ppeakerror}, the $y$-axis represents the relative deviation 
in the estimated $\ppeak$ from the subregions, and the $x$-axis represents the 
confidence in the estimated $\ppeak$. It is clear that the blueish data points for 
$\ell/\lf \lesssim 1$, either have a large relative error or a large range in the 
95\,percentile interval in the estimated $\ppeak$. For subregions of size $\ell/\lf < 1$, 
determining $\ppeak$ is subject to large uncertainties ($\gtrsim20\%$) mostly 
originating from stochastic variations between the subregions, or low $\ncc(>\pf)$, 
especially for $\kf\,L/2\uppi = 2$. Hence, even though the overall uncertainties and 
the 95\,percentile range in $\ppeak$ are mostly within 20\% of the $\pmean$, 
stochastic variations could impact/bias the results for subregions that have size 
$\ell \lesssim \lf$. On the other hand, for $\ell > \lf$, both stochastic variation and 
95\,percentile confidence in $\ppeak$ are on an average $\lesssim15\%$ level,
and our results can be safely employed.

Note that, for frequencies $\lesssim 1$\,GHz, since $\ncc(>\pf)$ is generally
larger compared to those at high frequencies (see Figure~\ref{fig:number_conn}), 
$\ppeak$ from the subregions could be determined to have better than $10\%$ 
accuracy for all $\kf$s for $\ell \approx \lf/2$.

\section{Conclusions}
\label{sec:conclusion}

Using synthetic $\pf$ maps of synchrotron emission originating due to the action of 
fluctuation dynamo, we showed that the number of connected polarized structures 
($\ncc$) and their filamentarity ($F$) is a powerful measure to understand the underlying 
driving mechanism in turbulent, magnetized plasma. $\ncc(>\pf)$ peaks at a certain 
$\ppeak$ which is directly related to the $\pmean$ of the entire emitting volume. 
This is significant because telescope noise in real observations makes the 
detection of faint polarized emission across the emitting medium difficult, which renders 
measuring $\pmean$ a challenging task. In such a case, $\ppeak$ can be used as a 
measure of $\pmean$ within $10\%$ accuracy. Since $\ppeak$ lies above $\sim0.1$, a 
typical level current telescopes can measure in diffuse media. For turbulence driving 
on scales ranging a factor of 4, $\ppeak$ can be measured using Minkowski functionals. 
For systems where turbulence is driven on $\lf<L/8$, $\ppeak \approx \pmean$ is 
expected at a lower value \citep[see][]{BS21} and likely will not be detectable. 
However, this will provide an upper limit on $\ppeak$.

Intrinsically larger polarized structures on scales comparable to $\lf$, that can be 
measured at high frequencies ($\gtrsim3\ghz$), tend to have larger $F$ compared 
to smaller structures, irrespective of their $\pf$. However, increasing Faraday 
depolarization toward low frequencies leads to the breaking up of these large-scale 
structures into smaller structures with $F$ higher than that of the intrinsic structure. 
This leads to more "canal-like" structures with $F\approx 1$, and can span over a 
significant fraction of the emitting medium. Furthermore, these canals span over 
scales that are comparable to the turbulence driving. For turbulence driven on small 
scales, the volume-filling polarized structures are more filamentary. Therefore, 
high-resolution observations are necessary to understand the origin of turbulence 
driving in subsonic media.

%%%%%%%%%%%%%%%%%%%%%%%%%%%%%%%%%%%%%%%%%%%%%%%%%%%%%%%%%%%%%%%%%%%
\begin{figure}
\hspace{-30pt}
\begin{centering}
\begin{tabular}{c}
{\mbox{\includegraphics[height=6.5cm, trim=0cm 0cm 0cm 0cm, clip]{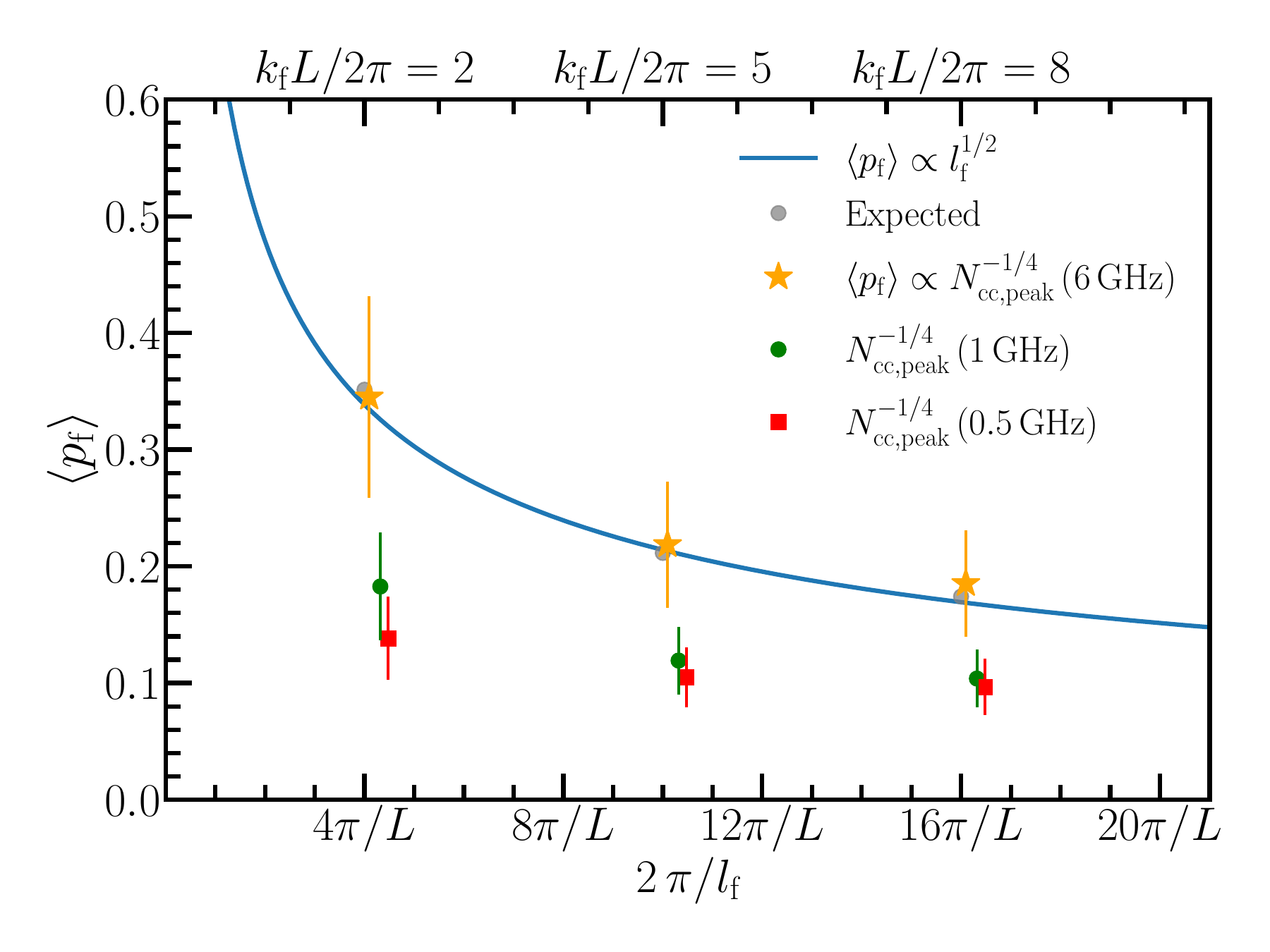}}} \\
\end{tabular}
\end{centering}
\caption{Variation of $\pmean$ with $\lf$. The gray points show the expected $\pmean$ 
computed from the magnetic integral scale ($\lmag$), and the blue curve shows the 
$\pmean \propto l_{\rm f}^{1/2}$ relation, both taken from \citet{BS21}. The red squares, 
green circles, and yellow stars show $\pmean \propto N_{\rm CC, peak}^{-1/4}$ at 
$0.5, 1$, and $6\ghz$ in the presence of Faraday rotation and are plotted with 
slight offset in the $x$-axis to avoid overlap. $N_{\rm CC, peak}$ are determined
from the entire emitting region of the simulated domain.}
\label{fig:Ncc_vs_k}
\end{figure}
%%%%%%%%%%%%%%%%%%%%%%%%%%%%%%%%%%%%%%%%%%%%%%%%%%%%%%%%%%%%%%%%%%%%%

Since $\ppeak$, as measured by $\ncc$ over the entire 2D map from the simulated
domain, is directly related to $\pmean$, and as shown in \citet{BS21}, at high frequencies, 
$\bra{p_{\rm f, \nu}} \approx \pmean_{\rm int}$ 
is related to $\lf$ as $\bra{p_{\rm f,\nu}} \propto \lf^{1/2}$, we expect $\ncc$ to be 
also important for determining the intrinsic fractional polarization ($\pmean_{\rm int}$) 
of an emitting medium. Figure~\ref{fig:Ncc_vs_k} shows the variation of $\pmean$ 
with $2\uppi/\lf$, wherein the different symbols show different estimators of $\pmean$. 
The blue solid curve shows the $\pmean \propto l_{\rm f}^{1/2}$ relation expected 
from a random walk of polarized emission along the LOS. The yellow stars, green 
circles and red squares show $N_{\rm CC,peak}^{-1/4}$ \textit{versus} $2\uppi/\lf$, 
where $N_{\rm CC, peak}$ corresponds to $\ncc(>\ppeak)$, and are determined at $6, 1$, 
and $0.5\ghz$, respectively. Interestingly, we find $N_{\rm CC, peak}^{-1/4}$ to be 
an excellent probe of $\pmean$ at higher frequencies, here $\nu = 6\ghz$, where it 
closely follows the $\pmean \propto l_{\rm f}^{1/2}$ relation. 
This alludes to an empirical relation $\lf \propto N_{\rm CC, peak}^{-1/2}$. 
This is not unexpected because the number of components is proportional to the 
area as $N_{\rm CC, peak} \propto (L/\lf)^2$. Thus, at high frequencies, the number 
of connected components per unit area of an emitting medium, $N_{\rm CC, peak}/L^2$, 
is a direct measure of the number of turbulent cells projected on the plane of the sky.

It is important to highlight that the magnetic integral scale ($\lmag$) varies linearly 
with $\lf$ \citep{BS21}. This implies that the variation of $\pmean$ shown in 
Figure~\ref{fig:Ncc_vs_k} can be equivalently represented in terms of $2\uppi/\lmag$. 
This will only change the quantitative values in the plot by a scale factor. However, 
for all practical purposes, comparisons with $\lf$ are more meaningful because of two 
reasons. Firstly, $\lf$ has a direct connection with the physical mechanism of turbulence 
driving. Secondly, while the value of $\lmag$ may depend on the Mach number of the 
flow and the $\Pm$, $\lf$ is a direct input in our simulations, which remains invariant 
throughout the evolution.

We note that, in order to infer $\lf$ using $N_{\rm CC, peak}$ discussed above, it is 
necessary that $\ppeak$ is well determined. To ensure that the detectable emitting 
region should be larger than $\sim\lf$. In the limiting case, when the region covers an 
area $\approx \lf\times\lf$, the inferred $\ppeak$ would have typical errors up to 
$\sim15\%$, which could arise due to a combination of stochastic variation and the 
accuracy to which $\ppeak$ can be estimated. However, because of the smaller area, 
the error on $N_{\rm CC, peak}$, $\delta N_{\rm cc,peak}/N_{\rm cc,peak}$, 
would roughly scale as $\sqrt{(L/\ell)^2}$, assuming Poisson distribution. A simple 
back-of-the-envelope calculation suggests that, in an extreme scenario, when 
$\ell \approx \lf$ and $\ell = \mathcal{O}(L/10)$, $\delta N_{\rm cc,peak}$ is 10 times 
larger. This would lead to up to a 50\% error in the estimated $\lf$. Although not very 
accurate, this would provide a realistic estimate of the driving mechanism in the 
media being studied.

Taken together, our results demonstrate that 2D Minkowski functionals, specifically, 
the surface density of the number of connected structures obtained from 
synthetic maps of $\pf$ for $\nu \gtrsim 3\ghz$, are a powerful tool to infer the 
underlying driving scale of turbulence.

\section{Discussion}
\label{s:discussion}

Our analysis of the Minkowski functionals and $\ncc$ relies on detectable 
polarized emission from a medium where fluctuation dynamo is believed
to operate, e.g., in the ICM and in the interstellar medium (ISM) of galaxies. In the 
context of the ICM, it is to be noted that there is a scarcity of detectable polarized 
emission in radio halos at present. This is due to the fact that polarized emission in
the ICM has been predominantly targeted near $1\ghz$. However, as shown in 
\citet{SBS21} and \citet{BS21}, high-frequency observations at $\nu \gtrsim 3\ghz$
are crucial to correctly infer the properties of polarized synchrotron emission arising 
from Fluctuation dynamo-generated fields. This is because, at these frequencies, 
the reduction in polarized emission from frequency-dependent Faraday depolarization 
and beam depolarization are significantly lower compared to that arising due to the 
steep spectrum of the synchrotron emitting cosmic ray electrons. With this in mind, 
our analysis of the morphological features at high frequencies presented in this work 
is a step toward the future, particularly in the light of next-generation radio telescopes 
when observations using MeerKAT between $2.6$ and $3.5\ghz$ and SKA1-MID 
between $2.8$ and $8.5\ghz$ would be a reality.

In the ISM of disk galaxies, it is important to note that both large-scale, mean 
magnetic fields generated by the galactic mean-field dynamo and the turbulent, 
small-scale magnetic fields generated by fluctuation dynamos coexist, 
arising from different, albeit related physical mechanisms \citep{beck16,SS21}. 
In addition to the small-scale fields, polarized emission also originates from large-scale 
fields on scales $L_{\rm mf} \gg \lf$, where $L_{\rm mf}$ is the scale of the galactic 
large-scale magnetic field. Thus, the direct use of $\ncc$ to infer $\lf$ in the local 
ISM requires the contribution from the large-scale fields to be isolated. Assuming 
that resulting polarized emission from large-scale fields is on scales $\ell_{\rm obs}$ 
such that $\lf < \ell_{\rm obs} < L_{\rm mf}$, the contribution from these fields can 
be removed by subtracting their nonzero $\bra{Q}$ and $\bra{U}$ determined over 
an area $\ell_{\rm obs}^2$. Nonetheless, we believe because inferring $\lf$ depends 
on $N_{\rm CC,peak}$, which will be dominated by components from turbulent 
fields, the presence of large-scale fields will not affect it significantly.

Furthermore, because the expected turbulence driving scale in the ISM is 
$\mathcal{O}(50\,\rm pc)$, physical resolutions of $\approx10\pc$ are needed. 
With the typical angular resolution of $\sim2^{\arcsec}\textrm{--}5^{\arcsec}$ 
available from current telescopes, galaxies at distance $\lesssim2$\,Mpc, 
e.g., M\,31, M\,33, LMC, IC\,10, etc., are the most suitable objects. For Galactic 
ISM, a careful consideration of diverging LOS with respect to observers on 
Earth is necessary. A detailed quantification is beyond the scope of this paper 
and will be addressed elsewhere.

\section*{acknowledgments}
R.D. thanks the Indian Institute of Astrophysics (IIA) for support under the 
Vacation Students Programme (VSP), where ideas of this work germinated. We 
thank Ancor Damas Segovia for the helpful comments and suggestions. 
Sharanya Sur acknowledges the use of the High Performance Computing 
resources made available by the Computer Centre of IIA. We thank the 
referee for comments that improved the discussion of the results in the paper. 
The software used in this work was developed in part by the DOE NNSA and 
DOE Office of Science supported Flash Center for Computational Science at 
the University of Chicago and the University of Rochester. This work made use 
of the {\sc SciPy} project \href{https://scipy.org}{https://scipy.org}, 
{\sc Numpy} \citep{2020NumPy-Array}, and {\sc Astropy}:\footnote{\href{http://www.astropy.org}{http://www.astropy.org}} 
a community-developed core Python package and an ecosystem of tools and 
resources for astronomy \citep{astropy22short}.

\newpage
\appendix

\section{Algorithm for computing Minkowski functionals} 
\label{sec:algo}

%%%%%%%%%%%%%%%%%%%%%%%%%%%%%%%%%%%%%%%%%%%%%%%%%%%%%%%%%%%%%%%%%%%%%%%%%
\begin{figure*}
\begin{centering}
\includegraphics[width=\textwidth]{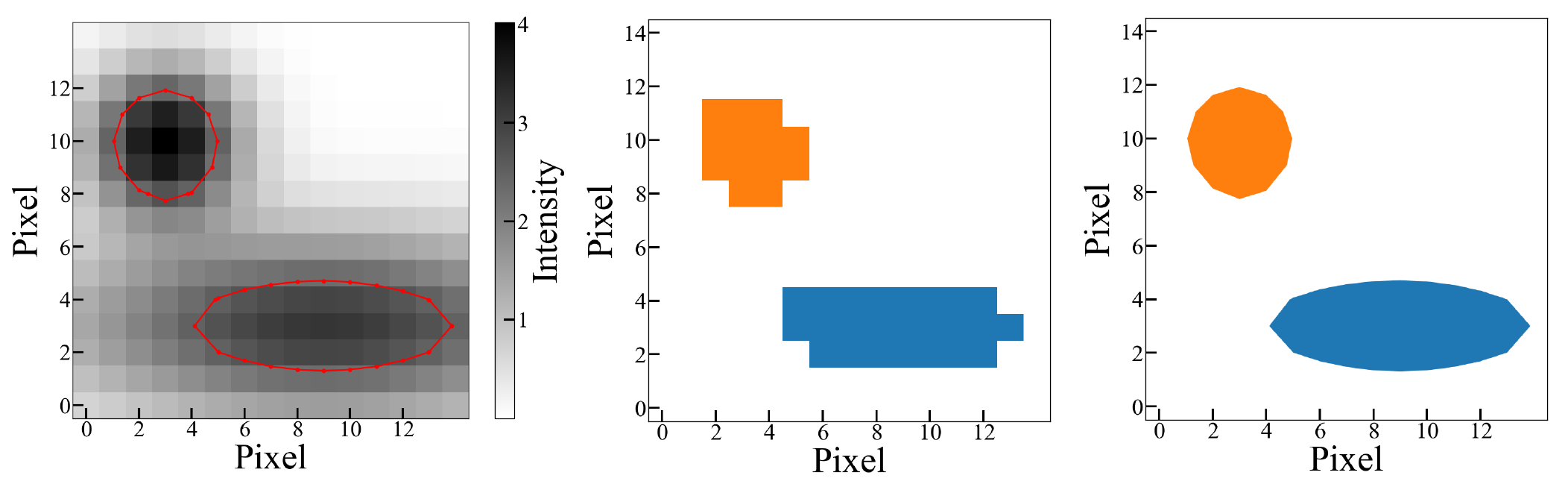}
\end{centering}
\caption{Illustration of the sequence of steps (left to right) performed 
by our algorithm for a gray-scale image formed by the superposition of 
two Gaussians. The threshold value of intensity is chosen to be $2.5$. 
The $(A,P,F)$ values obtained for the two connected components shown 
here are $(12.1,12.5,0.01)$ (yellow) and $(25.3,21.6,0.2)$ (blue).}
\label{fig:algo_flow}
\end{figure*}
%%%%%%%%%%%%%%%%%%%%%%%%%%%%%%%%%%%%%%%%%%%%%%%%%%%%%%%%%%%%%%%%%%%%%%%

We describe the new algorithm for computing the spatially resolved 
Minkowski functionals using the Python-based package 
\texttt{perimetrics} \citep{perimetricsv1_0_0}, developed for this work.
The two Minkowski functionals, area $A$ and perimeter $P$, of a 2D 
structure above a certain threshold value "$p$" are computed from the 
contour at level $p$. Since the contours of a gridded map are not smooth 
(limited by the pixelization) of an otherwise smooth 2D field, our 
modified algorithm first determines an approximately smooth contour 
via interpolation. The length of the contour and the area bounded 
within it give estimates of $P$ and $A$, respectively. The computation of 
$P$ and $A$ in our algorithm is based on the \textit{Marching squares} 
technique of \citet{mantz08}. As described in their work, we 
directly use the values of the map for determining appropriate 
contours instead of first converting them into binary maps as 
done in community-developed Python modules, e.g., \texttt{quantimpy} 
and \texttt{scikit-learn}. 

However, our implementation differs from \citet{mantz08} in the following 
key ways-- (i) finding and labeling the connected components so 
that in the final step, we can simply add the cell-/pixel-wise values 
of $A$ and $P$ separately for each connected component, and (ii) instead 
of computing $F$ from the global $A$ and $P$ as 
performed by \citet{mantz08}, we calculate them separately for 
each connected component. These have two important 
advantages. First, our approach reduces the computation complexity from 
$\mathcal{O}(N^2)$ to $\mathcal{O}(N)$, allows for a robust estimation 
of $F$, and retains information on individual connected components for 
spatially resolved analyses. The second can be readily seen from a 
simple example. Suppose that the structures in question are $N$ identical 
circles with radius $R$ that are isolated from each other, so that 
the total area $A = N\times\uppi\,R^2$ and the total perimeter, 
$P=N\times2\,\uppi\,R$. Then, Equation~(\ref{eq:filament_defn}) implies
$F = (N-1)/(N+1) \approx 1$ for large $N$. Thus, one would erroneously
conclude that an image that consists only of circles ($F=0$) to 
be highly filamentary. 

%%%%%%%%%%%%%%%%%%%%%%%%%%%%%%%%%%%%%%%%%%%%%%
\begin{figure*}[ht!]
\begin{centering}
\includegraphics[width=0.8\textwidth]{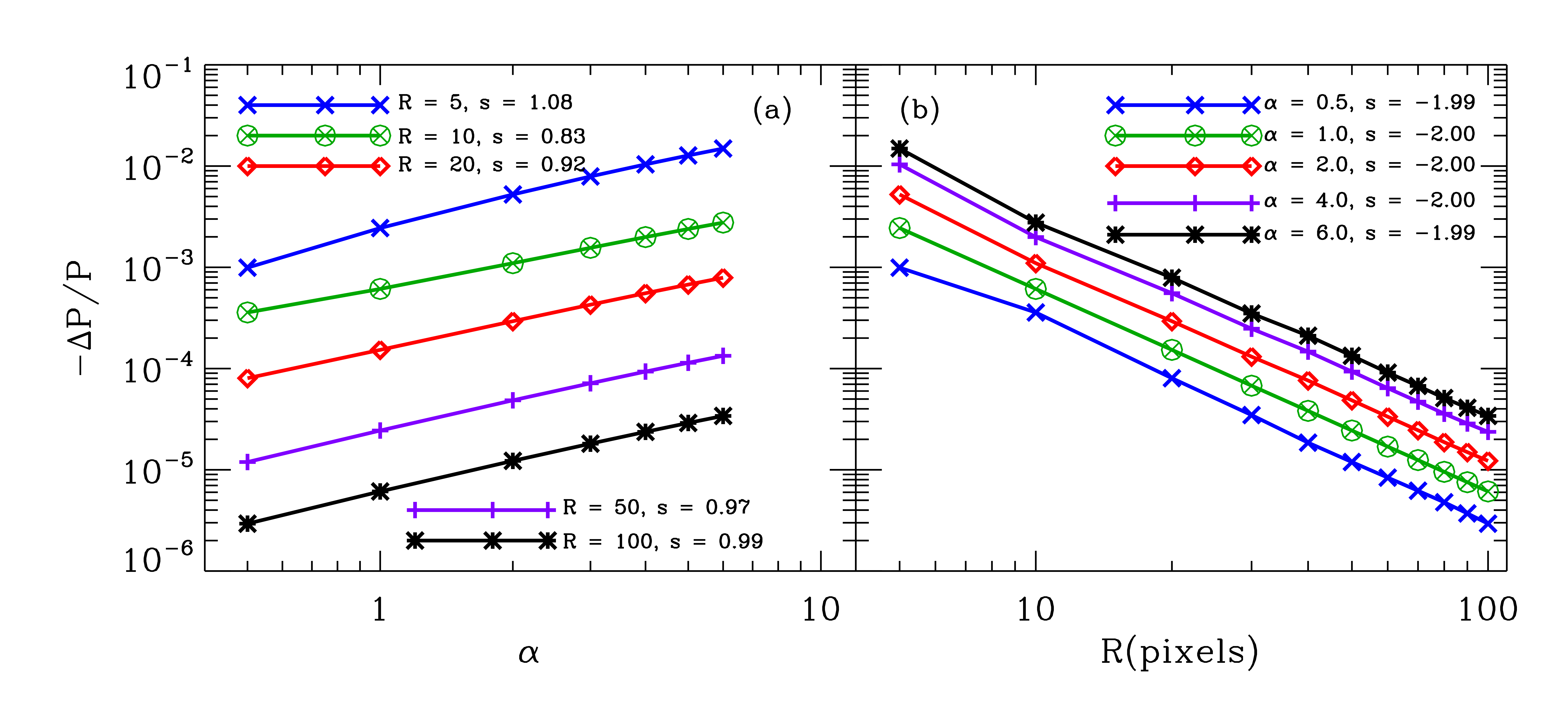}
\end{centering}
\caption{Plots for the fractional error in estimating the perimeter 
of a circular contour as a function of the exponent $\alpha$ of 
the power-law intensity profile (\textit{left}) and the radius $R$
of the circle (\textit{right}). The gradient of the intensity at the 
edge of the circle is given by $-\alpha/R$. The values of the slopes 
"$s$" corresponding to the best linear fits to the curves are shown in 
the legends.}
\label{fig:code_test}
\end{figure*}
%%%%%%%%%%%%%%%%%%%%%%%%%%%%%%%%%%%%%%%%%%%%%%%%%

We illustrate the working of our algorithm using a gray-scale image 
formed by the superposition of two Gaussians shown in the left 
panel of Figure~\ref{fig:algo_flow}. Each square 
pixel in this image is denoted as a "{\it cell}" and the 
corners of the cell as "{\it vertices}". The middle panel in the figure 
shows the colored labeling of the connected components, while 
the right panel shows the contours of these connected components. 
Our {\it Marching Square} algorithm performs 
three operations in sequence -- (a) segregating regions based on a 
threshold value and subsequently determining the points through which 
the contour should pass and assigning $A$ and $P$ values to 
each cell, (b) identifying and labeling the connected components 
using the $8$-connectivity criteria\footnote{Let "$S$" represent a 
subset of pixels in an image. Two pixels "$a$" and "$b$" are said 
to be {\it connected} in "$S$", if there exists a path between them 
consisting entirely of pixels in "$S$". In 2D, this can be obtained 
either using $4$- or $8$-connectivity. Thus, for any pixel "$a$" in "$S$", 
the set of pixels that are connected to it in "$S$" is called a 
{\it connected component} of $S$.}  \citep{GW02}, and (c) combining 
the component labels with cell-wise $A$ and $P$ values to quantify 
the area $A$ and the perimeter $P$ in each connected component and 
subsequently estimate their filamentarity ($F$). Note that step 
(a) measures the cell-wise values of $A$ and $P$ as performed by 
\citet{mantz08}, but it cannot distinguish between separate {\it structures}. 
Thus step (b) is essential to identify which pixel belongs to which {\it structure}. 
The terms 'threshold' and  'connected components' carry their usual meanings 
as used in image processing techniques \citep[see][]{GW02}. Further details 
can be accessed on the \texttt{Github} page of \texttt{permetrics}
\footnote{\url{https://github.com/rijudutta/perimetrics/tree/main}}.

\subsection{Test of the algorithm}
\label{sec:test_algo} 

%%%%%%%%%%%%%%%%%%%%%%%%%%%%%%%%%%%%%%%
\begin{figure}
\includegraphics[width=\columnwidth]{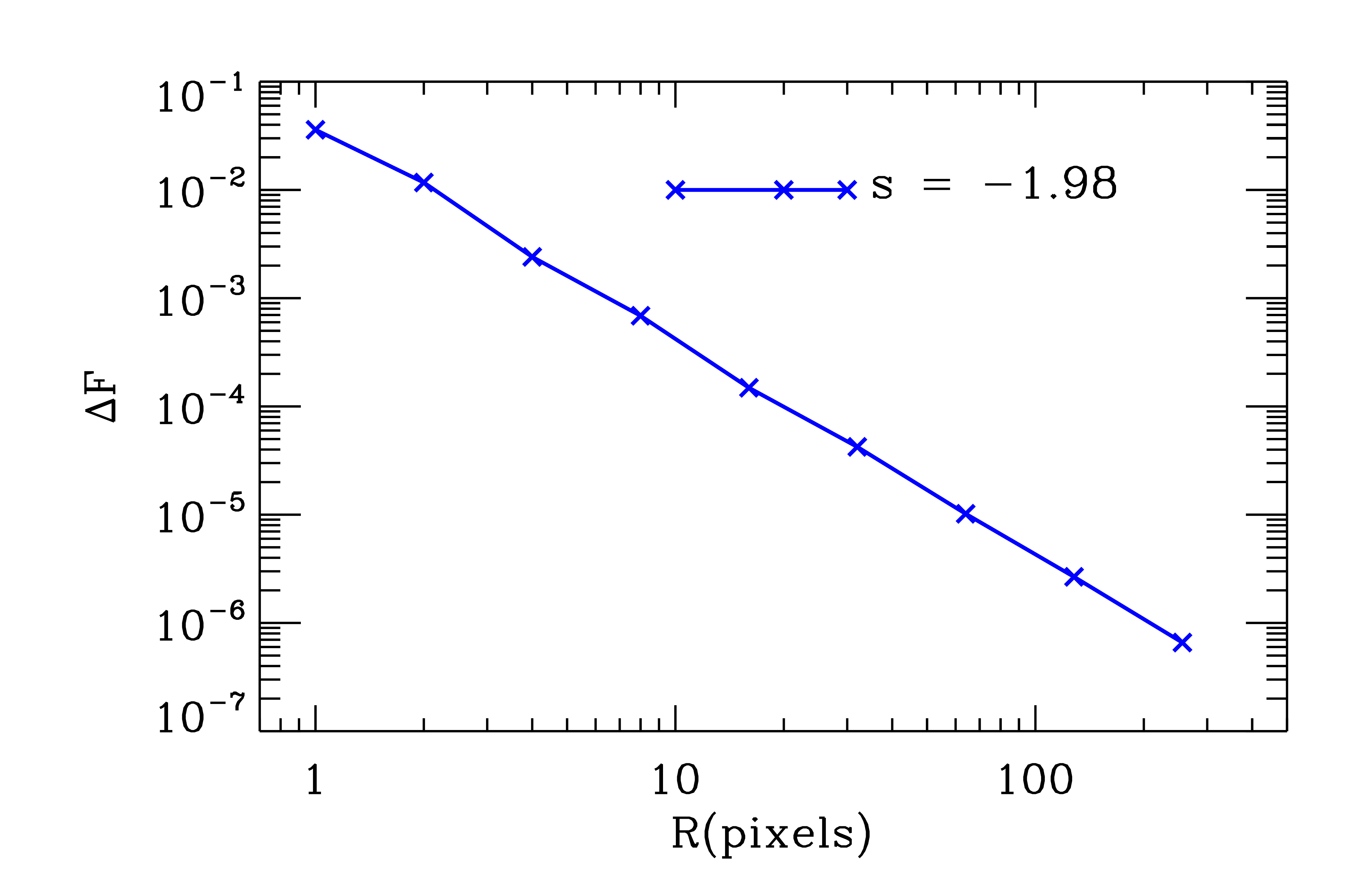}
\caption{Error in estimating the filamentarity ($\Delta F$) of a 
circular contour as a function of the radius $R$ of a circle for 
$\alpha=1$. Similar to Figure~\ref{fig:code_test}, the value of  
"$s$" corresponding to the best linear fit to the data points is 
shown in the legend.} 
\label{fig:F_error_test}
\end{figure}
%%%%%%%%%%%%%%%%%%%%%%%%%%%%%%%%%%%%%%%

To test the performance of the algorithm in determining $A$ and $P$ and their 
errors, we generate smooth, 2D intensity profiles given by,
\EQ
I(r) = 2 - \left(\frac{r}{R}\right)^\alpha.
\label{eq:test_profile}
\EN
Here $r$ is the radial distance from the center of the image, and the exponent 
$\alpha$ determines the gradient (steepness) of the profile at the characteristic 
radius $r=R$. Note that Equation~(\ref{eq:test_profile}) implies that the intensity 
profiles have circular symmetry, and therefore the structure being investigated 
have circular cross sections that always have $F=1$.

We set the intensity threshold at a value $I_{\rm thres} = 1$. This suggests that 
structures have radius $R$, and the gradient $\partial I/\partial r = -\alpha/R$ at 
the edge ($r = R$). The distances are measured in units of the pixel side length. 
Next, for a given $R$, we generate a $3R\times3R$\,pixel$^2$ image, and the 
origin, $r=0$, is defined at the center pixel. Then the connected component for 
our test cases forms a circle centered at the origin, which spans $R$ pixels between 
the origin and the edge of the circle. While "$A$" can be approximated quite well 
at higher resolutions by simply counting the number of cells inside a contour, 
accurate estimation of "$P$" is sensitive to the details of the contouring process. 
To this end, using our algorithm, we focus on computing $P$ and compare it with 
the expected value of $2\uppi R$.

Figure~\ref{fig:code_test} shows the dependence of the fractional error, 
$-\Delta P/P= -(P_{\rm comp} - 2\uppi R)/2\uppi R$, of the computed value of the 
perimeter $P_{\rm comp}$ on $\alpha$ (panel "(a)") and $R$ (panel "(b)"). We find 
that the error on $P$ decreases with both $\alpha$ and $R$ as a power law. By 
averaging over the exponents of the respective power laws for a range of $\alpha$ 
values evenly spaced between 1 and 6, and a range of $R$ values evenly spaced 
between 10 and 100, we report that the fractional error scales as $\alpha^{0.947} R^{-1.998}$. 
This suggests a scaling of $\alpha/R^2$ for the fractional error, which in turn implies 
that the absolute error in estimating the perimeter, $P - 2\uppi R$, is directly proportional 
to $-\alpha/R$, the gradient of the data at the radius of the contour. Thus, with this 
code, shallower gradients and higher resolution at the location of the contour lead 
to a more accurate estimation of the perimeter of the contour. 

Moreover, since $R$ is simply the number of pixels across the radius of the circle, the above 
results indicate that $P$ computed using our algorithm indeed converges to the expected 
$2 \uppi R$ as the resolution increases. It is also worth noting from Figure~\ref{fig:code_test} 
that even at a relatively low resolution, such as $R=5$ (with $\alpha=1$), the fractional error 
in estimating the perimeter to be $-2.4\times 10^{-3}$, while at $R=2$, it is $-1.7 \times 10^{-2}$. 
The corresponding fractional errors for area estimation are $-8.5 \times 10^{-3}$ and $-5.6 \times 10^{-2}$ 
respectively. For $\alpha=1$, the error in filamentarity $\Delta F$ (equal to the computed $F$, 
as the expected $F=0$ for a circle) is shown in Figure~\ref{fig:F_error_test}. We find that 
$\Delta F \sim R^{-2}$. Note that even for the extreme case of $R=1$, $\Delta F = 0.035$. 
Larger circles would have even smaller errors in $F$ as shown in our tests. When compared 
to mean values of filamentarity, $\meanF = 0.2 - 0.4$ found in our $\pf$ images, the average 
error can be expected to be $\lesssim 10\%$ of $\meanF$. 

Thus, the tests indicate that this algorithm can be reliably used to measure morphological 
quantities of turbulent structures spanning a wide range of sizes, including the smallest 
resolved structures. 

%\bibliography{minkowski_aa}{}
%\bibliographystyle{aasjournal}

\end{document}